\documentclass[twocolumn,preprintnumbers,amsmath,amssymb,longbibliography]{revtex4-2}

\usepackage{graphicx}
\usepackage{dcolumn}
\usepackage{float} 
\usepackage{bbm}

\usepackage{setspace}

\begin{document}

\title{Unusual Dynamical Properties of Disordered Polaritons in Microcavities}

\author{ Georg Engelhardt$^{1,2,3,4}$    }

\author{ Jianshu Cao$^{5} $}
\email{jianshu@mit.edu}

\affiliation{%
$^1$Beijing Computational Science Research Center, Beijing 100193, People's Republic of China.\\
$^2$Shenzhen Institute for Quantum Science and Engineering, Southern University of Science and Technology, Shenzhen 518055, China.\\
$^3$International Quantum Academy, Shenzhen 518048, China.\\
$^4$ Guangdong Provincial Key Laboratory of Quantum Science and Engineering, Southern University of Science and Technology, Shenzhen, 518055, China.\\
$^5$Department of Chemistry, Massachusetts Institute of Technology, 77 Massachusetts Avenue,
Cambridge, Massachusetts 02139, USA.
}

\date{\today}


\begin{abstract}
 	The collective light-matter interaction in microcavities gives rise to the intriguing phenomena of cavity-mediated transport that can potentially overcome the Anderson localization. Yet, an accurate theoretical treatment is challenging as the matter (e.g., molecules) is subject to large energetic disorder. In this article, we develop the Green’s function solution to the Fano-Anderson model and use the exact analytical solution to quantify the effects of energetic disorder on the spectral and dynamical properties in microcavities. Starting  from the microscopic equations of motion, we derive an effective non-Hermitian Hamiltonian and predict a set of scaling laws: (i) The complex eigen-energies of the effective Hamiltonian exhibit an exceptional point, which leads to underdamped coherent dynamics in the weak disorder regime, where the decay rate increases with disorder, and overdamped incoherent dynamics in the strong disorder regime, where the slow decay rate decreases with disorder. (ii) The total density of states of disordered ensembles can be exactly partitioned into the cavity, bright-state, and dark-state local density of states, which are determined by the complex eigensolutions and can be measured via spectroscopy. (iii) The cavity-mediated relaxation and transport dynamics are intimately related such that both the energy-resolved relaxation and transport rates are proportional to the cavity local density of states. The ratio of the disorder-averaged relaxation and transport rates equals the molecule number, which can be interpreted as a result of a quantum random walk. (iv) A turnover in the rates as a function of disorder or molecule density can be explained in terms of the overlap of the disorder distribution function and the cavity local density of states. These findings reveal the significant impact of the dark states on the local density of states and consequently their crucial role in optimizing spectroscopic and transport properties of disordered ensembles in cavities.
\end{abstract}

\maketitle

\allowdisplaybreaks

\section{Introduction.}

The control of polaritons in microcavities has seen  rapid experimental and theoretical progress in recent years, in particular, in the field of molecular polaritons~\cite{Garcia-Vidal2021}. Experiments have revealed intriguing phenomena such as  large Rabi splittings~\cite{Weisbuch1992,Shalabney2015,George2016} and  enhanced transport properties in  organic semiconductors~\cite{Lerario2017}. The intriguing physics originates from the  collective light-matter interaction, which is induced by  the strong  photon confinement in the cavity. This leads to an effective Rabi splitting proportional to  the square root of the total number of quantum emitters (e.g.,  molecules as specified in this article) $\sqrt{N}$. The large Rabi splitting is a consequence of the formation of a collective  bright state, which couples coherently  to the light field, resulting in two polaritonic states. Yet, the overwhelming number of states are decoupled from the light field. These states are denoted as 'dark states' and 'dark-state reservoir' in the literature.
%
%
%

Bright and dark states are theoretically well-understood for homogeneous systems without disorder. In the presence of disorder, the nature of these states is qualitatively understood in terms of photon borrowing, i.e., a mixing of the dark states and the cavity mode mediated by  disorder~\cite{Gonzalez-Ballestero2016,Sommer2021,Houdre1996,Lopez2007,Spano2015,Shammah2017,Herrera2017a,Xiang2019}. While the physical properties of polaritons have been characterized in the weak disorder regime~\cite{Litinskaya2006,Agranovich2003,Litinskaya2004}, the nature of the bright and dark states for arbitrary disorder has not been  investigated quantitatively and rigorously. 

 For systems with only local couplings, disorder gives rise to  Anderson localization, which is particularly prominent in low-dimensional systems~\cite{Anderson1958}. For a one-dimensional system, an infinitesimal amount of disorder induces a localization of the wave function, which results in an exponentially suppressed conductivity. 
 %
 %
 It is well known that the coupling of a quantum system to a thermal bath gives rise to environment-assisted transport, which can help to overcome the localization of an excitation or charge  such that the energy mismatch between different sites can be compensated by a noise-induced level broadening of the local site energies~\cite{Cao2009}.  This mechanism is relevant in excitation transport in light-harvesting systems and molecular semiconductors~\cite{Wu2013,Chin2010,Lambert2013,Rebentrost2009,Chuang2016,Lee2015,Moix2013a}. 
 In the latter case, charge mobility or exciton diffusion shows a turnover as a function of the system-environment coupling, where quantum transport is enhanced by spatial coherence at small couplings but exponentially suppressed by dynamic localization (i.e., the polaron effect) at strong couplings. 
 In contrast,  static disorder has predominately a detrimental effect on the transport properties even in the presence of long-range hopping ~\cite{Rodriguez2003}. The disorder-assisted transport in cavities studied here is a rare exception. 
 %
%

For matter interacting with the electromagnetic field in a microcavity, enhanced ~\cite{Hou2020,Krainova2020,Orgiu2015} and unaffected transport efficiency~\cite{Rozenman2018} for molecular excitations and charges has been observed depending on the experimental setup. Theoretical studies using numerical and analytical methods have also been reported~\cite{Cao2022,Chavez2021,Feist2015,Schachenmayer2015,Hagenmueller2017}.
However, a thorough theoretical understanding of the cavity-mediated transport properties  based on simple concepts is lacking, as the theoretical treatment of  disorder is  challenging.

\begin{figure}
	\includegraphics[width=1\linewidth]{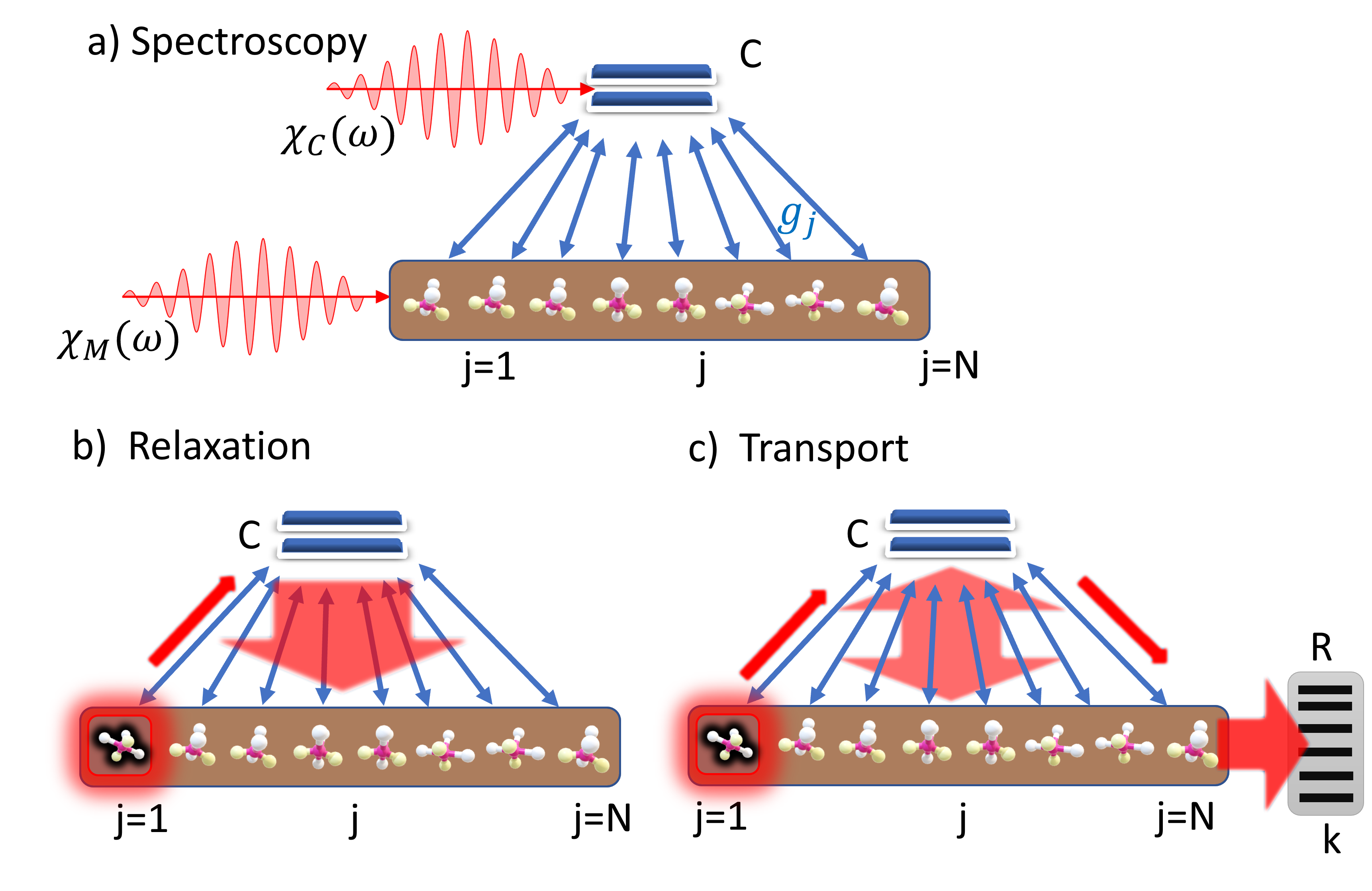}
	\caption{Sketches of the system which consists of  $N$ molecules (or atoms etc.) labeled by $j$, which are coupled to a single cavity mode (C). The blue arrows depict the tree-like coupling structure of the model. (a) depicts  two spectroscopic experiments which measure the cavity absorption $\chi_{C}(\omega)$ and the matter absorption $\chi_{M}(\omega)$. (b) depicts a relaxation process, where an initial excitation on the donor  $j=1$ spreads over all molecules as illustrated by the red arrows. (c) depicts a  transport process, where an initial excitation on the donor  is  transported to the reservoir (R), which is coupled to the acceptor molecule $j=N$. As illustrated by the red arrows, the excitation will spread over the molecules first before being finally  transported to the reservoir. }
	\label{figSketchCavity}
\end{figure}

This article has two major objectives. First, we introduce a nonperturbative theoretical framework to investigate  polariton dynamics using the Green's functions in Laplace space. 
Second, we use this framework to establish a simple and unified picture describing the spectroscopic, relaxation and transport properties of  disordered ensembles.

\textbf{Theoretical framework.}  We employ   Green's functions based on the equations of motion  and generalize methods developed in Ref.~\cite{Engelhardt2016a,Topp2015}.  The treatment in the Laplace space enables a  microscopic derivation of an effective  Hamiltonian describing the cavity mode and the bright state. Its non-Hermitian structure incorporates the effect of the dark states, and its  complex eigenvalues give  insight into the polariton dynamics. We consider a Lorentzian energetic disorder, which allows for compact expressions for spectroscopic and transport properties.
Moreover, we  develop two  analytical methods: (i) The first method is  the polynomial perturbation theory (PPT), which unifies the standard degenerate and non-degenerate perturbation theories. This unified perturbation theory is thus suitable for systems with a continuous energy spectrum as considered here. (ii) The second method is the exact stochastic mapping (ESM). It maps one system configuration to another, which has the same stochastical properties but a more convenient structure for  further analytical calculations. 

\textbf{Physical picture.}  The central quantity of this unified physical picture is the cavity local density of states (LDOS), which reveals the mixing of bright and dark states in the presence of disorder and can be measured by cavity absorption as shown in Fig.~\ref{figSketchCavity}(a).  The linewidth of the cavity LDOS exhibits a turnover:  The width first increases with small disorder as more dark states become coupled to the cavity field, and then decreases with large disorder as more dark states move out of resonance with the cavity field.  

The second half of the article explores the effects of disorder on the cavity-mediated
relaxation and transport processes for the experimental setups sketched in Figs.~\ref{figSketchCavity}(b) and~\ref{figSketchCavity}(c), respectively. As a key result, both the energy-resolved relaxation rate and the resonant  transport rate can be expressed in terms of the cavity LDOS.  The two processes are intimately related as the ratio of the disorder-averaged relaxation rate and  transport rates exactly equals the number of molecules, which can be interpreted in terms of a quantum random walk.  Interestingly, the rate can be optimized as a function of disorder or molecular density, which is a consequence of  the overlap of the cavity LDOS and the disorder distribution. This type of turnover has been observed as a function of dephasing rate in noise-assisted quantum transport, where disorder usually suppresses coherence and transport~\cite{Wu2013,Kassal2013,Ishizaki2012,Lambert2013,Scholes2014,Rebentrost2009,Chuang2016,Lee2015,Moix2013a}. Now, due to the collective coupling to the cavity field, the disorder-assisted transport can also exhibit the intriguing turnover behavior.

\textbf{Layout.} This article is organized as follows. In Sec.~\ref{sec:SystemMethods}, we explain the system and introduce the Green's function based on the exact equations of motion. In Sec.~\ref{sec:SingleParticleObservables}, we investigate the LDOS of different  constituents and relate them to spectroscopic properties as sketched in Fig.~\ref{figSketchCavity}(a).  In  Sec.~\ref{sec:relaxtionDynamics}, we investigate  the relaxation dynamics sketched in Fig.~\ref{figSketchCavity}(b). In Sec.~\ref{sec:excitationTransfer}, we analyze the  transport process sketched in Fig.~\ref{figSketchCavity}(c). In Sec.~\ref{sec:discussion}, we summarize the results and provide future research perspectives.  In the Appendixes, we provide  detailed derivations. In particular, the PPT and the ESM, which are applied in the calculations of the relaxation and transport properties, are explained in detail in Appendix~\ref{sec:AnalyticalTechniquesDetails}.

\section{System and methods}

\label{sec:SystemMethods}

\subsection{Hamiltonian}

We consider a  microcavity   containing  $N$ quantum emitters labeled by $j$ as sketched in Fig.~\ref{figSketchCavity}. To enable analytical calculations, we describe the electromagnetic field in the cavity with a single mode labeled by ($C$). The Hamiltonian describing the system reads as
\begin{equation}
\hat H = \hat H_{L} + \hat H_{M} +  \hat H_{LM},
\label{eq:lightMatterHamiltonian}
\end{equation}
where the light, matter, and light-matter interaction Hamiltonians are given as
\begin{eqnarray}
\hat H_{L} &=& E_C  \hat a^\dagger \hat a ,   \nonumber  \\   
\hat H_{M} &=& \sum_{j=1}^{N } E_j \hat B_{j}^\dagger \hat B_j  ,\nonumber  \\
\hat H_{LM} &=&  \sum_{j=1}^{N }  g_j  \hat a \hat B_{j}^\dagger + \text{H.c.} ,  
\label{eq:hamiltonianTerms}
\end{eqnarray}
respectively.
 The cavity mode  is quantized by the photonic operator $\hat a $ and has energy $E_C$.  The two-level systems, which are described by the operators $\hat B_j$, refer to general quantum emitters, such as atoms,  charges, excitons, spins, electronic or vibrational  levels of molecules.  In the following, we focus on molecular excitations because of their experimental relevance, but our findings are generally valid for the other before mentioned systems.
  The  excitation energies $E_j$ are distributed according to the Lorentz function
\begin{eqnarray}
P(E_j) = \frac{1}{\pi}\frac {\sigma} { \left(E_j -E_M \right)^2 + \sigma^2}, 
\label{eq:LorentzDistribution}
\end{eqnarray}
where $E_M$ is the center of the probability distribution and $\sigma$ is its width, i.e., the disorder parameter. We emphasize that many of our results  hold for arbitrary disorder distributions.  The light-matter couplings can be expressed in terms of  physical quantities as $g_j = \left( \frac{\hbar E_C}{2 \epsilon_0 V } \right)^{(1/2)} \mathbf{d}_j \cdot \mathbf{\hat E} $, where $V$ is the volume of the cavity, $\mathbf{d}_j$ is the dipole moment of the $j$-th molecule, and $\mathbf{\hat E}$ is the polarization  of the cavity mode. For simplicity, we  assume here a homogeneous coupling $g_j=g$, but generalizations to the inhomogeneous case are straightforward. Many calculations are performed in a thermodynamic limit which is defined as $g\rightarrow 0$ and $N\rightarrow \infty$ such that $g\sqrt{N} $ is constant. As $g \propto V^{-1/2}$, the thermodynamic limit implies a large  cavity volume, but a constant molecule density.

 The Hamiltonian in Eq.~\eqref{eq:lightMatterHamiltonian} is the celebrated Fano-Anderson model, which has been originally developed to understand the impact of continua on discrete levels and asymmetries in absorption spectra~\cite{Fano1961}. Aside from other applications, this and generalized Fano-Anderson models are also deployed to investigate transport through nanoscopic and mesoscopic systems~\cite{Nitzan2003,Engelhardt2019a,Boehling2018,Gallego-Marcos2017}. A multi-mode version of this model has been numerically studied in Ref.~\cite{Ribeiro2021}. A recent investigation of the spectral and transport properties  for a  disorder distribution with compact support can be found in Refs.~\cite{Dubail2021,Botzung2020}, which make use of the exact expression of the eigenstates instead of  the unifying  Green's function approach considered here. We note that the transport setup investigated in Refs.~\cite{Dubail2021} is different from the one considered in our work in Sec.~\ref{sec:excitationTransfer}.

\subsection{Bright and dark states in the homogeneous system}

\label{sec:darkStatesBrightStates}

The homogeneous system is defined for  vanishing disorder $\sigma=0$. For simplicity, we focus here on the resonant system  $E_M =E_C$. Using the excited states of the molecules  $\left| e_j \right>_M   = \hat B_j^\dagger \left| g \right>$, where $\left| g \right>$  is the collective ground state of the molecule ensemble, we define the collective excitations $\left| e_k \right>_M   = \frac{1}{  \sqrt{N} } \sum_j e^{i k j  }  \left| e_j \right>_M $ with $k = 2\pi l/N $ and $l= 0, ...,N-1$. The Fock states of the cavity mode are denoted by $\left| n\right>_L$. In terms of these states, the two special eigenstates of the Hamiltonian 
\begin{equation}
    \left|\psi_{up/down} \right>	 = \frac{1}{\sqrt{2} }\left( \left| g \right>_M \left| 1\right>_L \pm  \left| e_{k=0} \right>_M  \left| 0\right>_L  \right),
\end{equation}  
with energies $\epsilon_{0,1} =  E_M \pm g \sqrt{N}$ are called the upper and lower polaritons, respectively. Their energy difference $\Omega_R= 2  g \sqrt{N}$, which increases with the square root of the molecule number, can be measured as a collective Rabi splitting $\Omega_R$ in spectroscopic experiments. The homogeneous state $\left| BS \right>_M = \left| e_{k=0} \right>_M $, which mixes with the cavity light field, is commonly denoted as the bright state. In contrast, the states $\left| DS_k \right>_M  = \left| e_{k\neq0} \right>_M $, which have energy $\epsilon_k = E_M$, completely decouple from the cavity light field and are  denoted as dark states. We note that the bright and dark states can be also defined for an inhomogeneous coupling $g_j$. In this case, the bright state reads as $\left| BS \right>_M   \propto \sum_j g_j \left| e_j \right>_M $, and the dark states are orthogonal to this state.

For the disordered system, the dark states are no longer eigenstates of the system, but all eigenstates have contributions of the cavity mode, the bright state and the dark states, as the matrix elements $\left <BS \right|\hat H_M \left| DS_k\right>\neq 0$ of the matter Hamiltonian in Eq.~\eqref{eq:hamiltonianTerms} become finite.
The respective contributions of the cavity mode, the bright state and the dark states to the eigenstate is given respectively by the  bright-state LDOS and the dark-state LDOS, which will be introduced in  Sec.~\ref{sec:SingleParticleObservables}. The mixing of bright and dark states influences the spectroscopic spectra in Sec.~\ref{sec:SingleParticleObservables} and leads to the turnover in the relaxation and transport rates in Secs.~\ref{sec:relaxtionDynamics} and \ref{sec:excitationTransfer}.

\subsection{Single-molecule Green's function}

\begin{figure}
	\includegraphics[width=0.99\linewidth]{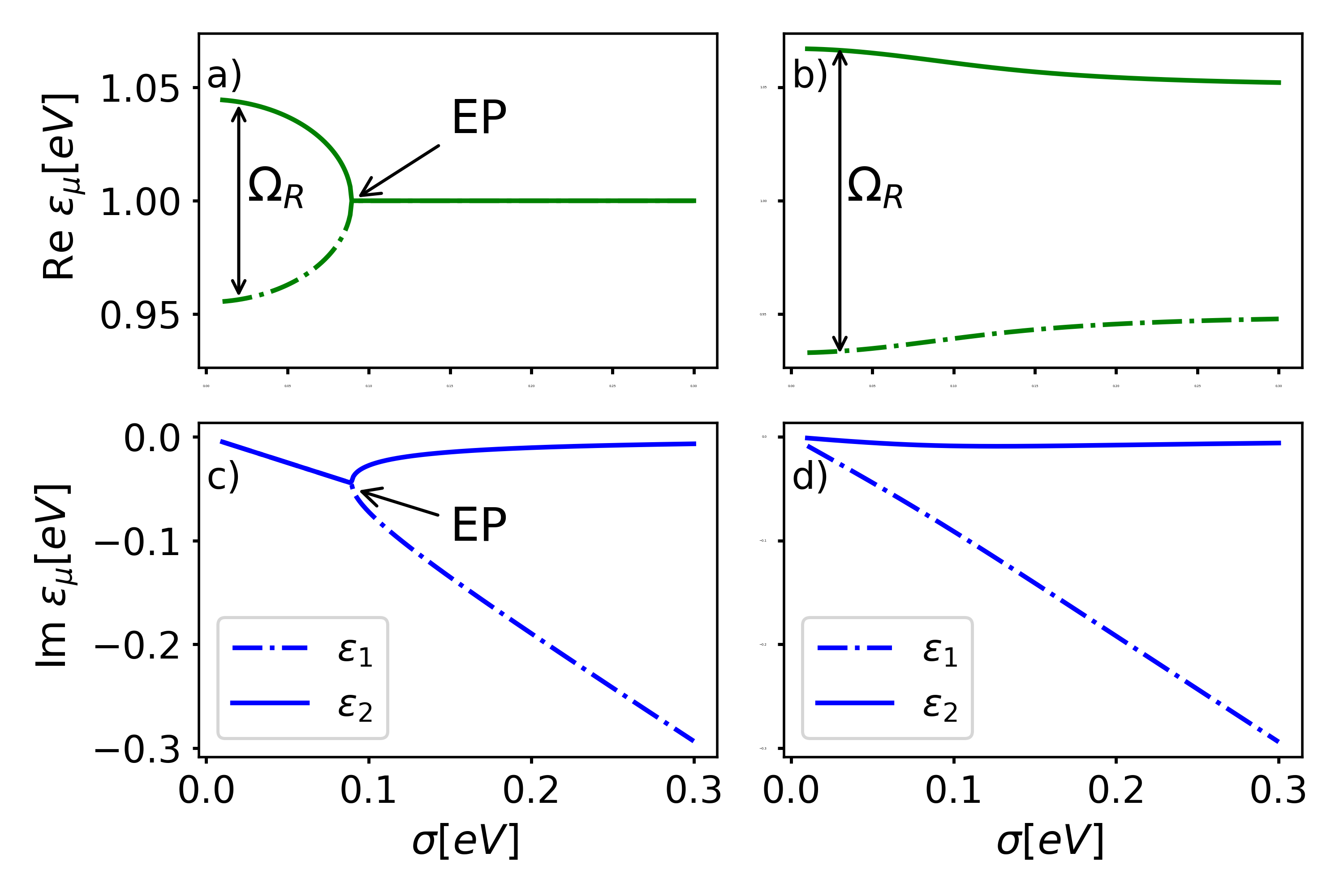}
	\caption{(a), (b) Real and (c), (d) imaginary parts of the eigenenergies in Eq.~\eqref{eq:RootsCharPolynomial} of the effective non-Hermitian Hamiltonian  for $g=0.001E_M$  and  $N=2000$. (a), (c) depict the resonant system with $E_C= E_M=1\;\text{eV}$ and (b), (d) depict the off-resonant system with $E_C=1.05\;\text{eV}$ and $ E_M=0.95\;\text{eV}$.   }
	\label{figRootsNoDephasing}
\end{figure}

As shown in Fig.~\ref{figSketchCavity}, the Fano-Anderson model has a tree-shaped coupling structure, which can be solved analytically in Laplace space as shown in Appendix~\ref{sec:DetailsSolutionEoMLaplace}. In doing so, we find an explicit expression of the single-particle retarded Green's functions 
\begin{equation}
  G_{X,Y} (t) \equiv i \Theta(t)\left< \hat A_X (t), \hat A_Y^\dagger\right>,
  \label{eq:defRetardedGreensFunction}
\end{equation}
where $ X \in \left\lbrace C,j ,BS, DS_k \right\rbrace$,  and $\Theta(t)$ denotes the Heaviside step  function. Thereby, $\hat A_C^\dagger = \hat a^\dagger$, $\hat A_j^\dagger = \hat B_j^\dagger$ and $\hat A_{BS}^\dagger  $ ($\hat A_{DS_k}^\dagger  $ ) creates a bright state  (dark state) excitation.  In Laplace space, the Green's functions read as~\cite{Engelhardt2016a,Topp2015}
\begin{eqnarray}
G_{C,C} (z) &=&   
\frac{ 1 }{ \left( z + i E_C(z) \right) } , \nonumber   \\
G_{C,j} (z) &=&   
  -i \frac{ g  }{\left( z + i  E_C (z) \right) \left( z + i E_j \right) } = G_{j,C} (z)  \nonumber  , \\
	G_{i,j} (z) &=&   
	  \frac{ \delta_{i,j} }{z + iE_j }  - \frac{ g^2  }{\left( z + i E_i \right)\left( z + i  E_C (z) \right) \left( z + i E_j \right) } \nonumber ,\\
	\label{eq:greensFktDD}
\end{eqnarray}
with the  auxiliary function 
\begin{eqnarray}
E_C (z) &=&  E_C  -i  \sum_{j=1}^{N} \frac{g^2}{z + i E_j  } .
\label{eq:auxilaryFunction}
\end{eqnarray}
Up to factors $(z+iE_j)$,  the nominator of the third term in Eq.~\eqref{eq:greensFktDD} defines a polynomial $\mathcal P(z)$ of order $N+1$,
\begin{equation}
\mathcal P(z) =   \left(  z + i  E_C (z)  \right)  \prod_j^{N} (z + iE_j),
\label{eq:characteristicPolynomial}
\end{equation}
which  is equivalent to the  characteristic polynomial of the Hamiltonian \eqref{eq:lightMatterHamiltonian} when replacing $z\rightarrow - i E$. The inverse Laplace transformation can be expressed in terms of  the roots of the characteristic polynomial, i.e., the poles of the Green's function, as
\begin{equation}
G_{X,Y} (t) = \sum_{\alpha=1}^{N +1} A_\alpha e^{z_\alpha t },
\label{eq:inverseLaplaceTranformation}
\end{equation}
where $A_\alpha = 2\pi i \lim\limits_{z\rightarrow z_\alpha} (z-z_\alpha ) G_{X,Y} (z)$.  Note that the pole of the first  term in $G_{j,j} (z)$ of Eq.~\eqref{eq:greensFktDD} is not a pole of $G_{j,j} (z)$, as it cancels with a pole of the second term.

\subsection{Thermodynamic limit and effective Hamiltonian}

\label{sec:effectiveHamiltonian}

\begin{figure*}
	\includegraphics[width=1\linewidth]{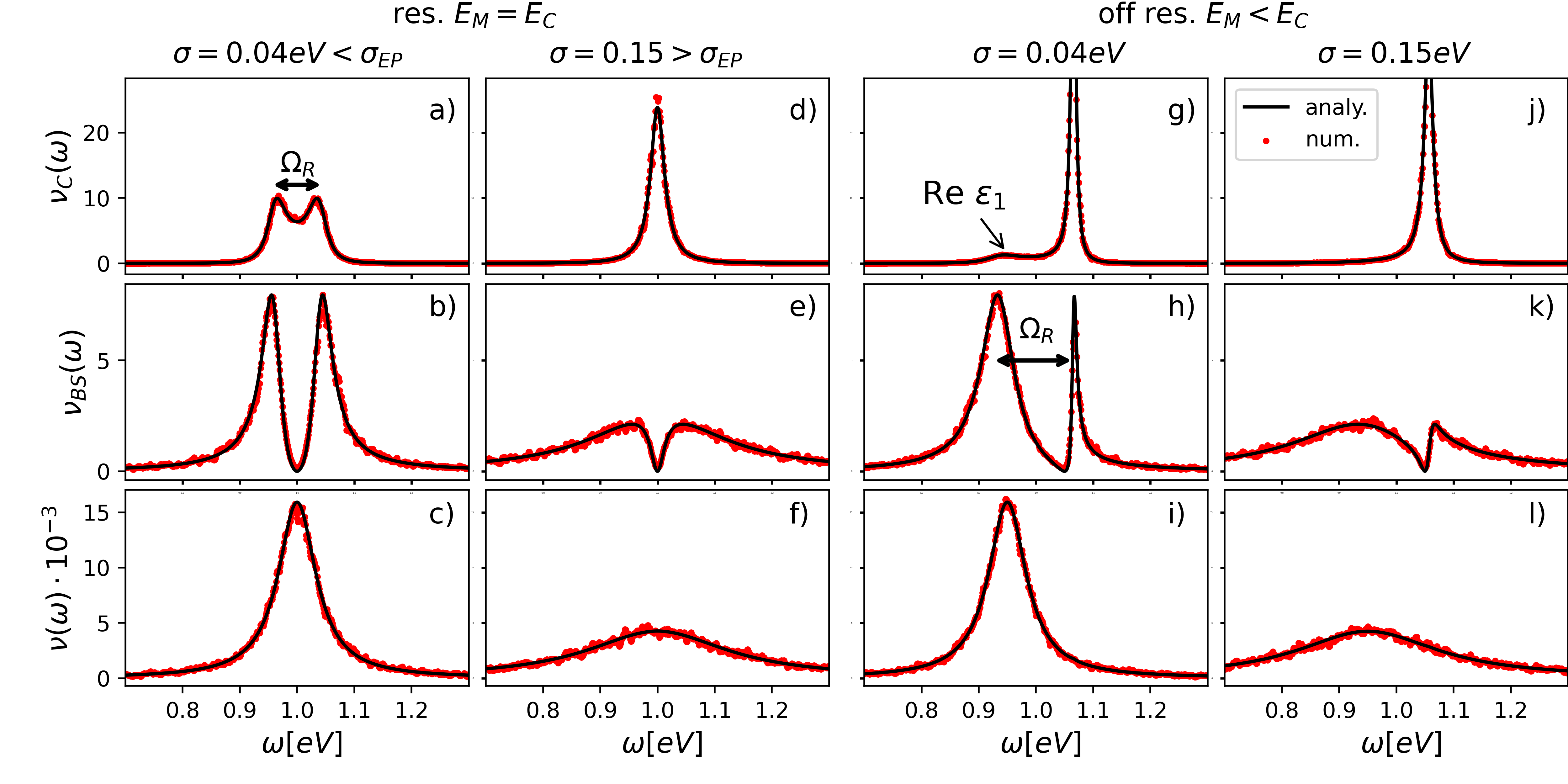}
	\caption{Analytical and numerical calculations of the cavity LDOS (a,d,g,j), the bright-state LDOS (b,e,h,k), and the total density of states (c,f,i,l) as a function of energy. Overall  system parameters are $g=0.001\;\text{eV}$ and $N=2000$. Specific  parameters are $E_C = E_M =1\;\text{eV}$, $\sigma =0.04\;\text{eV}< \sigma_{EP}$ in (a-c), $E_C = E_M =1\;\text{eV}$, $\sigma =0.15\;\text{eV}> \sigma_{EP}$,  $E_C =1.05\;\text{eV},  E_M =0.95\;\text{eV}$, $\sigma =0.04\;\text{eV}< \sigma_{EP}$ in (d-e), $E_C =1.05\;\text{eV},  E_M =0.95\;\text{eV}$, $\sigma =0.04\;\text{eV}$ in (g-h) and  $E_C =1.05\;\text{eV},  E_M =0.95\;\text{eV}$, $\sigma =0.15\;\text{eV}$ in (j-l).  The LDOS in all panels are depicted in units of $1/\text{eV}$. }
	\label{figEigenStateAnalysis}
\end{figure*}

Here, we consider the system in the thermodynamic limit   $N\rightarrow \infty$ and $g\rightarrow 0$ such that $g^2 N = \text{const}$, and derive an effective Hamiltonian which exactly reproduces the dynamics of the single-particle Green's function.
 The disorder-averaged Green's function  in the thermodynamic limit is defined by
\begin{equation}
	\overline G_{X,Y} (z) \equiv \int   dE_1 ... \int   dE_N G_{X,Y} (z) P(E_1)...P(E_N),
	\label{eq:disorderAverage}
\end{equation}
where $P(E_j)$ is the disorder distribution function of  $E_j$.
When evaluating $\overline G_{X,Y} (z)$, $ \sum_{j=1}^{N} \frac{g^2}{z + i E_j  } \rightarrow \Pi(z) $ in Eq.~\eqref{eq:auxilaryFunction} becomes a smooth function which depends on the statistics of $E_j $. Moreover, $\overline G_{i_1,j_1} (z) = \overline G_{i_2,j_2} (z)$ and $\overline G_{C,j_1} (z) = \overline G_{C,j_2} (z)$ for all $i_1,i_2,j_1,j_2$, i.e., the disorder-averaged Green's function is  homogeneous. As a consequence,  the Green's function is  block-diagonal  in the basis of  bright and dark states introduced in Sec.~\ref{sec:darkStatesBrightStates}, i.e., 
$\overline G_{e_{k_1},e_{k_2}}  (z) \propto \delta_{k_1,k_2}  $ and $\overline G_{C,e_{k}}  (z) \propto  \delta_{k_,0}  $. We use the disorder-averaged Green's function to define an effective Hamiltonian by
\begin{equation}
	\overline G (z) \equiv \frac{1}{z +  i H_{eff} (z) }.
\end{equation}
Note that the effective Hamiltonian depends on $z$ and the distribution of  $E_j$.  The block structure of  $\overline G (z) $ in the basis of bright and dark states translates into a block structure of $H_{eff} (z)$, i.e., 
\newcommand\bigzero{\makebox(0,0){\text{\Large 0}}}
\begin{equation}
	 H_{eff}(z) = 
	 \left[
	 \begin{array}{cccc}
	 	H_{eff} ^{(C,BS)}(z) & & & \bigzero \\
	 	& H_{eff} ^{(DS)}(z)  & & \\
	 	 & & \ddots & \\
	 	 \bigzero & & & H_{eff} ^{(DS)} (z)
	 \end{array}
	 \right],
	 \label{eq:effectiveHamiltonian}
\end{equation}
where $	H_{eff} ^{(C,BS)}(z) \in \mathbb C^{2\times 2}$ describes the interaction of the cavity mode and the bright state and $	 H_{eff} ^{(DS)} (z) \in \mathbb C$ describes the dynamics of the dark states. As the effective dark state Hamiltonians are equal for all $k\neq 0$, we have suppressed the index $k$.  The effective Hamiltonian Eq.~\eqref{eq:effectiveHamiltonian} appears to suggest that the dynamics of the bright  and  dark states is decoupled. Yet, the influence  of the dark states is incorporated in the non-Hermitian nature of $H_{eff}(z)$, which effectively leads to a dissipative dynamics.

\subsection{Effective Hamiltonian for the Lorentz distribution}

For the Lorentz distribution in Eq.~\eqref{eq:LorentzDistribution}, the evaluation of the disorder-averaged Green's function is straightforward and is equivalent to simply replacing  $E_j \rightarrow E_M -i \sigma \equiv E_M^{(\sigma)}$~\cite{Litinskaya2019,Herrera2021}. Because of this simple replacement rule, the thermodynamic limit considered in Eq.~\eqref{eq:disorderAverage} is actually equivalent to the disorder average, such that the following results are also valid for finite molecule numbers. The disorder-averaged Green's functions  are explicitly given  as
\begin{eqnarray}
\overline G_{C,C} (z) &=&   
\frac{ 1 }{ \left( z + i E_C(z) \right) } , \label{eq:greensFktDDdisorderAveraged}  \\
\overline G_{C,j} (z) &=&   
-i \frac{ g  }{\left( z + i  E_C (z) \right) \left( z + i E_M^{(\sigma)} \right) } = \overline G_{j,C} (z)  \nonumber ,  \\
\overline G_{i,j} (z) &=&   
\frac{ \delta_{i,j} }{z + iE_M^{(\sigma)} }   \nonumber  \\
&-&  \frac{ g^2  }{\left( z + i E_M^{(\sigma)} \right)\left( z + i  E_C (z) \right) \left( z + i E_M^{(\sigma)} \right) },
\nonumber
\end{eqnarray}
where now $ E_C (z)  = E_C + g^2N /( z + E_M^{(\sigma)}) $. Transforming this into the basis of the bright and dark states, the blocks of the effective Hamiltonian in Eq.~\eqref{eq:effectiveHamiltonian} are given as
\begin{equation}
	H_{eff} ^{(C,BS)} = 
	\left[
	\begin{array}{cc}
	E_C  & \frac\Omega 2  \\
	 \frac\Omega 2   & E_M^{(\sigma)}
	\end{array}
	\right]
	\label{eq:effectiveHamiltonianCBS}
\end{equation}
with the Rabi frequency of the homogeneous system $\Omega =2 g\sqrt{N}$ and  $ H_{eff} ^{(DS)} = E_M^{(\sigma)} $. Note that the  $z$-independence of the effective Hamiltonian is a consequence of the Lorentz distribution of  $E_j$.
 The complex-valued eigenenergies of  $H_{eff} ^{(C,BS)}$ in Eq.~\eqref{eq:effectiveHamiltonianCBS}
\begin{eqnarray}
\epsilon_{1,2} &=&\frac{E_C + E_M^{(\sigma)} }{2}\nonumber \\
& \pm&  \frac{1}{2}  \sqrt{\left(E_C -  E_M^{(\sigma)} \right)^ 2 -  \Omega^2  }
\label{eq:RootsCharPolynomial}
\end{eqnarray}
 generalize the eigenenergies of  the two polaritons for the disordered case.  As it will become more clear when discussing spectral properties, the real part is related to the spectral position and the imaginary part to the spectral width of the absorption line shape. The Rabi splitting of the disordered system can be thus defined as $\Omega_R \equiv\text{Re} \left(\epsilon_2 - \epsilon_1\right)$. The finite imaginary part appears due to the mixing  of the bright and dark states, which eventually gives rise to a decaying occupation of the subsystem consisting of cavity mode and bright state.

\textbf{Resonant system.} 
For the resonant system  $E_C =  E_M$, the eigenenergies are given as
\begin{eqnarray}
\epsilon_{1,2} &=&E_M - i\frac \sigma 2   \pm  \frac{1}{2}  \left(\Omega^2 -\sigma^2 \right)^{\frac 12}
\label{eq:RootsCharPolynomialResonance}
\end{eqnarray}
and are depicted in Figs.~\ref{figRootsNoDephasing}(a) and~\ref{figRootsNoDephasing}(c).
For very small or very large $\sigma$, they approximately read as
\begin{eqnarray}
\sigma \ll \Omega:  \qquad	\epsilon_\mu&=& E_M -i \frac{\sigma}{2} \pm \left( \frac\Omega 2 - \frac{\sigma^2}{\Omega}\right) ,\\
\sigma \gg \Omega:  \qquad \epsilon_\mu &=& 
\begin{cases}
E_M -i\sigma + i \frac{\Omega^2}{2\sigma}  & \mu =1\\
E_M - i \frac{\Omega^2}{4\sigma} & \mu =2.
\end{cases}
\label{eq:asymptoticBehaviorEnergies}
\end{eqnarray}
These two limiting cases can be clearly seen in  Figs.~\ref{figRootsNoDephasing}(a) and~\ref{figRootsNoDephasing}(c). For small $\sigma$, the two roots have different real parts, and  equal imaginary parts. For large $\sigma$, the real parts are equal, but the imaginary parts differ. The agreement of either the real or the imaginary part is a consequence of $E_C =E_M$.
Interestingly, for $\sigma=  \Omega  \equiv \sigma_{EP}\approx 0.09 eV $, we observe an exceptional point, which has been frequently investigated  recently~\cite{Miri2019}.  In Secs.~\ref{sec:CavityLocalDensityOfStates} and~\ref{sec:absorptionSpectrum}, we discuss  signatures of the exceptional point in the absorption spectrum. In the following, we denote the region before and after the exceptional point as the underdamped  (small $\sigma$) and overdamped  (large $\sigma$) regime, respectively.

\textbf{Off-resonant system.} For the off-resonant system with $E_M < E_C$ in Figs.~\ref{figRootsNoDephasing}(b) and~\ref{figRootsNoDephasing}(d), the real and imaginary parts  of both eigenvalues are well separated for all values of $\sigma$. Even for a small detuning of $E_M $ and $ E_C$, the exceptional point observed in the resonant system does not exist. The eigenvalue $\epsilon_1$  is dominated by the molecular excitations, while $\epsilon_2$  is  dominated by the cavity excitation. The imaginary part (describing the spectral width) of $\epsilon_2$ is much smaller than the imaginary part of $\epsilon_1$, as the light and matter become increasingly decoupled for larger disorder due to the decreasing density at $E_j \approx E_C$ in the off-resonant system. 

\section{Local density of states and spectroscopic properties}

\label{sec:SingleParticleObservables}

Spectroscopic and transport observables can be expressed in terms of the LDOS associated with  system constituents $X\in   \left\lbrace C,j,BS,DS_k\right\rbrace$.  These LDOS are defined  in terms of the diagonal elements of the single-particle retarded Green's function
\begin{eqnarray}
\nu_{X}  (\omega)= - \lim_{\delta\rightarrow  0^+}\text{Im} \frac{1}{\pi} G_{X,X} (-i\omega+ \delta) ,%
\label{eq:localDensityOfStates}
\end{eqnarray}
which quantifies how much a specific system state $X$ contributes to the eigenstates in the energy interval $\left[ \omega ,\omega + d\omega \right]$.  The total density of states is given by $\nu(\omega) = \sum_{X \in \left\lbrace C, j \right\rbrace } \nu_{X}  (\omega)$. As we explain in the following, the cavity and bright-state LDOS can be measured with spectroscopic experiments.

\subsection{Cavity absorption spectrum }

\label{sec:CavityLocalDensityOfStates}

First, we investigate the spectroscopic response of the cavity related to the perturbation operator $\hat V =  \hat a + \hat a^\dagger $, which will be denoted as the cavity absorption spectrum in the following. In terms of the Green's function, the cavity absorption can be expressed as
\begin{eqnarray}
\chi_C (\omega) &=& - \text{Im} \; G_{C,C} \left( -i\omega +0^+ \right) 
=  \pi \nu_C (\omega)
\label{eq:CavityLocalDensityOfStatesDef}
\end{eqnarray}
and is thus directly proportional to the cavity LDOS. For the Lorentz distribution in the thermodynamic limit, we evaluate Eq.~\eqref{eq:CavityLocalDensityOfStatesDef} using the disorder-averaged Green's function $\overline G_{CC}(z)$  given in Eq.~\eqref{eq:greensFktDDdisorderAveraged}, which can be transformed into
\begin{eqnarray}
\chi_C (\omega) &=&  \sum_{\mu =1,2 }  \frac{-1}{\pi}\text{Im } \left[  \frac{A_\mu}{\omega - \epsilon_\mu } \right] \nonumber, \\
A_\mu &=&  \frac {-i \epsilon_\mu +  i E_M +\sigma}{-i \epsilon_\mu + i \epsilon_{\overline \mu }},
\label{eq:CavityLocalDensityOfStates}
\end{eqnarray}
where the coefficients $A_\mu$ can be evaluated in terms of the energies $\epsilon_\mu $ in Eq.~\eqref{eq:RootsCharPolynomial}. Details of the derivation can be found in Appendix~\ref{sec:detailsCavityAbsoporptionSpectrum}. We have introduced  $\overline\mu$  as $ \overline\mu \neq \mu$ for a compact notation.
If  $A_\mu$ is   real-valued, the cavity absorption spectrum is given by two Lorentz functions centered at $\text{Re}\, \epsilon_\mu$ and having spectral width  $\text{Im} \,\epsilon_\mu$. For complex-valued $A_\mu$, the shape   deviates from the pure Lorentzian function. As the finite imaginary part of the eigenenergies describes the mixing of  the bright and dark states, the dark states thus determine the functional form of the cavity LDOS. 

In Figs.~\ref{figEigenStateAnalysis}(a) and~\ref{figEigenStateAnalysis}(d), we depict the cavity absorption spectrum for the resonant case $E_M=E_C$    in the underdamped and overdamped regimes, respectively. In the underdamped regime, we observe two  Lorentzian peaks symmetrically located around $E_M$. Here, the eigenenergies fulfill $ (- \epsilon_1 +  E_M ) = ( \epsilon_2 - i E_M )^*$  as can be seen in Fig.~\ref{figRootsNoDephasing}, which explains the symmetry of the peaks when evaluating Eq.~\eqref{eq:CavityLocalDensityOfStates}.
In the overdamped regime, we observe a single peak with Lorentzian shape. Here, the eigenenergies fulfill $ \epsilon_1 +  i \sigma/2 =  \epsilon_2 - i \sigma/2$, such that there are actually two Lorentzian peaks centered at the same position $E_M$. The eigenenergy $\epsilon_2$ with a small imaginary part  dominates the spectrum. Using Eq.~\eqref{eq:asymptoticBehaviorEnergies} to evaluate $A_1$, we find that $A_1\rightarrow 0$ for $\sigma\rightarrow \infty$, such that the signatures of the eigenenergy $\epsilon_1$ are strongly suppressed. A physical interpretation of the eigenenergies in the overdamped regime is  given in the next section.

Figures~\ref{figEigenStateAnalysis}(g) and~\ref{figEigenStateAnalysis}(j)   depict the cavity absorption spectrum for the off-resonant system $E_M< E_C$    in the underdamped and overdamped regimes, respectively.  Similar to the resonant system, we observe a peak close  the position of the cavity frequency $E_C \approx \text{Re}\;\epsilon_{2}$. In  Fig.~\ref{figEigenStateAnalysis}(g), we mark  a second peak related to the molecular eigenenergy $\epsilon_1$ located close to $E_M$, which is very small as the corresponding $A_1\rightarrow 0$ for large $\left| E_C -E_M\right|$.  The observations in the overdampend regime in Fig.~\ref{figEigenStateAnalysis}(j) are very similar, but the molecule peak is now completely suppressed.

\subsection{Matter absorption spectrum}

\label{sec:absorptionSpectrum}

Next, we  consider the absorption spectrum which is related to the perturbation operator $\hat V =\sum_j D_j \left( \hat B_j + \hat B_j^\dagger \right)$, which we denote as matter absorption in the following. The coupling coefficients between the molecules and the probe field are  $D_j  =  \mathbf{d}_j \cdot \mathbf{\hat E}_p$, where  $\mathbf{\hat E}_p$ is the probe field. It  measures directly the spectroscopic properties of the molecules instead of the cavity field as in  Sec.~\ref{sec:CavityLocalDensityOfStates}.
Assuming a homogeneous $D_j =D$, we can express the matter absorption spectrum in terms of the single-particle Green's functions as
\begin{eqnarray}
\chi_M (\omega) &=& - D^2 \sum_{i,j=1}^{N}  \text{Im }  G_{i,j} (-i\omega+0^+ ) \nonumber \\
&=& N D^2 \pi \nu_{BS} (\omega) .
\label{eq:molecularAbsorptionGreensFunktDef}
\end{eqnarray}
In contrast to the absorption spectrum of uncorrelated molecules, where only the diagonal elements of the Green's function are taken into account, the absorption of the molecular polaritons includes all elements of the Greens' function as the cavity induces strong coherences between the molecules. 

The summations in Eq.~\eqref{eq:molecularAbsorptionGreensFunktDef} project  the Green's function onto the bright state introduced in Sec.~\ref{sec:darkStatesBrightStates}, such that the matter absorption spectrum is directly proportional to the bright-state  LDOS. We note that the relation of the matter absorption and the bright-state density of states in Eq.~\eqref{eq:molecularAbsorptionGreensFunktDef} is also correct for inhomogeneous couplings as long as $g_j \propto D_j$, i.e., the polarizations of the cavity and the probe field are parallel.

For the Lorentz distribution in the thermodynamic limit,  we can use the  disorder-averaged Green's function  in  Eq.~\eqref{eq:greensFktDDdisorderAveraged} to evaluate the matter absorption, such that we  find after some steps
\begin{eqnarray}
\chi_M (\omega) &=&  \sum_{\mu =1,2 }  \frac{-D^2}{\pi}\text{Im } \left[  \frac{A_\mu}{\omega - \epsilon_\mu } \right] \nonumber, \\
A_\mu &=&  \frac {g^2 }{\left( - \epsilon_\mu+  i E_M +\sigma\right) \left( -i\epsilon_\mu + i \epsilon_{\overline \mu } \right) }.
\label{eq:molecularAbsorptionGreensFunkt}
\end{eqnarray}
 As explicitly demonstrated in Appendix~\ref{sec:molecularAbsorptionSpectrum}, the matter absorption spectrum can be expressed in terms of the cavity single-particle Green's function as
\begin{eqnarray}
\chi_M (\omega) &=&  N D^2\text{Im} \;G_{CC} \left( i\omega + iE_M + iE_C +\sigma \right) ,
\label{eq:molecularAbsorption}
\end{eqnarray}
i.e., using a complex frequency  $\omega \rightarrow -\omega + E_M + E_C + i\sigma $. Thus, the absorption spectra of the cavity mode and of the matter are directly related to each other.

The molecular absorption is depicted in Figs.~\ref{figEigenStateAnalysis}(b) and~\ref{figEigenStateAnalysis}(e) for the resonant system $E_M = E_C$. Because of  Eq.~\eqref{eq:molecularAbsorption},  $\chi_M (\omega) $ has the same functional dependence as the cavity absorption Eq.~\eqref{eq:CavityLocalDensityOfStates} except for the complex-valued frequency. The discussions about the cavity absorption are thus also valid for the matter absorption. In the underdamped regime in Fig.~\ref{figEigenStateAnalysis}(b), we find two peaks corresponding to the two polaritons which approximately have a  Lorentzian shape, similar to the cavity absorption in  Fig.~\ref{figEigenStateAnalysis}(a).  Interestingly, in Figs.~\ref{figEigenStateAnalysis}(b),~\ref{figEigenStateAnalysis}(e),~\ref{figEigenStateAnalysis}(h), and~\ref{figEigenStateAnalysis}(k) the matter absorption vanishes  completely at $\omega =  E_M$ due to level repulsion of the molecules which are in resonance with $E_C$. In contrast, the level repulsion for the cavity LDOS in Figs.~\ref{figEigenStateAnalysis}(a),~\ref{figEigenStateAnalysis}(d),~\ref{figEigenStateAnalysis}(g), and~\ref{figEigenStateAnalysis}(j) is not complete as the cavity mode is interacting with a continuous spectrum which leads to a coarse graining of the level repulsion.

%
%
 The complete suppression of the absorption in Fig.~3 is reminiscent of the electromagnetically-induced transparency~\cite{Fleischhauer2005} (EIT) and the related vacuum-induced transparency (VIT)~\cite{Field1993,Litinskaya2019} appearing in atomic and molecular three-level systems. In fact, the absorption suppression in Figs.~\ref{figEigenStateAnalysis}(b),~\ref{figEigenStateAnalysis}(e),~\ref{figEigenStateAnalysis}(h), and~\ref{figEigenStateAnalysis}(k)  can be also understood as a destructive interference between two excited states. However,  in contrast to the EIT and VIT, which considers  individual three-level atoms or molecules coupled by a light field, in this paper, the light field itself is also a quantum state and couples the molecules collectively. Moreover, the suppressed absorption observed here  is based on a  non-local superposition of the quantum emitters  (associated with the bright state) and is thus a collective effect, while the  EIT and VIT are based on a destructive interference effect within individual atoms or molecules.

 In the overdamped regime depicted in Figs.~\ref{figEigenStateAnalysis}(e) and ~\ref{figEigenStateAnalysis}(k), the matter absorption strongly deviates from the cavity absorption despite their close relationship via Eq.~\eqref{eq:molecularAbsorption}. In contrast to the cavity absorption,  both eigenenergies depicted in Fig.~\ref{figRootsNoDephasing} are now relevant. The matter absorption is a superposition of two Lorentz functions which have  different widths $\text{Im} \,\epsilon_1$ and $\text{Im} \,\epsilon_2$, but the same peak position of $\omega= E_M=E_C$. One peak has a positive amplitude, while the other has a negative amplitude, leading to a complete destructive interference at the cavity energy $E_C$. This complete destructive interference is  a consequence of the level repulsion of the cavity mode and the molecular excitation energies with $E_j = E_C$.  While the two peaks  in the underdamped regime can be associated with the two polaritons, this picture breaks down in the overdamped regime. The distinct behavior can be understood in terms  of the imaginary parts in Fig.~\ref{figRootsNoDephasing}. For an increasing disorder $\sigma$, the molecular excitations and the  cavity mode gradually decouple. The spectral features of the molecular excitations thus approach the original Lorentz distribution with width $\sigma$, while the spectral width of the cavity continuously vanishes. In this regime, the  polaritonic excitations are not well defined.
 
 The off-resonant system $ E_M <  E_C  $ in the underdamped regime depicted in Fig.~\ref{figEigenStateAnalysis} (h) is not symmetric. The molecule peak close to $\omega = E_M$ is significantly broader than the cavity peak close to $\omega = E_C$, which is directly related to the $\text{Im} \;\epsilon_{1,2}$ in Fig.~\ref{figRootsNoDephasing}. The cavity  LDOS is thus only weakly influenced by the energetic disorder because of the energy splitting. The observations in the overdamped regime in Fig.~\ref{figEigenStateAnalysis}(k) are similar to Fig.~\ref{figEigenStateAnalysis}(e) but not symmetric.

\subsection{Density of states}

\label{sec:densityOfStates}

Finally, we consider the total density of states  to clarify the role of the bright and dark states  introduced in Sec.~\ref{sec:darkStatesBrightStates} in the presence of disorder. As the total density of states does not depend on the basis, it can be expressed either in the local basis of the molecules $j$ or  in the basis of bright and darks states, 
\begin{eqnarray}
\nu  (\omega) &\equiv&   \nu_{C}  (\omega) + \sum_j \nu_j (\omega) \nonumber \\
&=& \nu_{C}  (\omega) + \nu_{BS}  (\omega) +  \sum_{k=1}^{N-1} \nu_{DS_k}(\omega)   ,
\label{eq:systemDensityOfStates}
\end{eqnarray}
  i.e, it can be partitioned in the cavity LDOS, $\nu_{C}  (\omega)$, the bright-state LDOS, $\nu_{BS}  (\omega)$, and $N-1$ terms of the dark-state LDOS. In the thermodynamic limit, the cavity and bright-state LDOS converge to their respective limits given in Sec.~\ref{sec:CavityLocalDensityOfStates} and Sec.~\ref{sec:absorptionSpectrum}, while the dark-state LDOS  converge to $\nu_{DS_k}(\omega)  \rightarrow (N-1)  P(\omega)$. 
   Thus, from the LDOS of the molecules in the non-interacting case $g=0$, which is  $ N P(\omega) $, one molecules is subtracted, which now forms the bright-state LDOS $\nu_{BS}  (\omega)$. 

In  the homogeneous and resonant  system discussed in Sec.~\ref{sec:darkStatesBrightStates}, we have $ \nu_{C}  (\omega) = \nu_{BS}  (\omega) = \frac{1}{2} \left( \delta(\omega - E_M + g \sqrt{N } ) +  \delta(\omega -E_M -g \sqrt{N } )  \right) $ and $\nu_{DS}  (\omega) = (N-1) \delta(\omega -E_M)$. Comparing these expressions with the LDOS considered in this section, we find that the disorder turns the delta functions into distribution functions with  finite widths. In the homogeneous case the molecular excitations can be either classified as bright or dark states. Because of the disorder,  bright and dark states are now mixed and both contribute to the formation of the eigenstates. On the average, the contributions are thereby proportional to $\nu_{BS}(\omega)$ or $\nu_{DS}(\omega)$, respectively. We recall that  the dark states determine the functional shape of $\nu_{C}(\omega)$ and $\nu_{BS}(\omega)$, as  the dark states are coupled to the bright state for a finite disorder.

In realistic spectroscopic experiments, one simultaneously measures  the cavity and matter absorption, i.e., $\chi(\omega) = \alpha_C \cdot  \chi_C(\omega) + \alpha_M \cdot  \chi_M(\omega)$ with  coefficients $\alpha_C$ and $\alpha_M$. For a fixed Rabi frequency $\Omega^2 =4 g^2N \propto N/V$, one can thus harness the scaling $\chi_{C} \propto \nu_{C}$ and $\chi_{M} \propto N \cdot \nu_{BS} \propto V\cdot \nu_{BS}$ to distinguish experimentally between both contributions via changing the cavity volume $V$ while keeping the molecule density constant.

\section{Relaxation dynamics}

\label{sec:relaxtionDynamics}

In this section, we evaluate the relaxation dynamics of an excitation, which is initially located on the  donor molecule $j=1$ and finally spreads over all molecules. The process is sketched in Fig.~\ref{figSketchCavity}(b).  In the thermodynamic limit, the donor  will be finally completely depleted. 
As we  demonstrate in the following, the relaxation dynamics of the cavity system is proportional to the cavity LDOS considered in Sec.~\ref{sec:CavityLocalDensityOfStates}. In Sec.~\ref{sec:derivationRelaxationRate}, we derive the relaxation rate. Readers  interested in the physical interpretation can proceed directly  to Sec.~\ref{sec:discussionRelaxationRate}.

\subsection{Derivation of the relaxation rate}

\label{sec:derivationRelaxationRate}

To begin with, the occupation of the donor  is given as
\begin{equation}
n_1(t) =  \left| G_{1,1} (t ) \right|^2 ,
\end{equation}
where the Green's function is defined in Eq.~\eqref{eq:greensFktDD}. To facilitate an analytical treatment, we apply the disorder average in the thermodynamic limit defined in Eq.~\eqref{eq:disorderAverage}. Yet, to account correctly for the microscopic dynamics of the donor occupation, the average is not applied  to $E_1$. Specifying  for the Lorentzian disorder, the Green's function can be written in Laplace space as
\begin{eqnarray}
\overline G_{1,1} (z ) 
&=&  \frac{ 1  }{z + iE_1 } - \frac{ g^2 \left( z + i E_M+\sigma  \right) }{\left( z + i E_1 \right) \mathcal P(z)   }  
\label{eq:disorderdAveragedGreensFunctionRelaxation}.
\end{eqnarray}
Thereby, the roots of the polynomial $\mathcal P(z) = \mathcal P_0(z)+\mathcal P_1(z)$ with
\begin{eqnarray}
\mathcal P_0(z) &=&  \left( z + i E_C \right)\left( z + i E_M +\sigma \right) \left( z + i E_1 \right) \nonumber,  \\
 &&+g^2N  \left( z+i E_1 \right), \nonumber  \\
\mathcal P_1(z)  &=&         g^2\left( z_{\alpha} + i E_M +\sigma \right)\nonumber 
\end{eqnarray}
determine the  inverse Laplace transformation in Eq.~\eqref{eq:inverseLaplaceTranformation}. Note that $z=-iE_1$ is not a root of $\overline G_{1,1} (z )$ as the corresponding terms cancel exactly.
The third-order  polynomial is partitioned into the two  parts $\mathcal P_0(z)$ and $\mathcal P_1(z)$. The first part can be solved analytically and we obtain the unperturbed  roots  $z_1^{(0)}= -i \epsilon_1$ and $z_2^{(0)}= -i \epsilon_2$ and $z_3^{(0)} = -i E_1$, with $\epsilon_{1,2}$ given in  Eq.~\eqref{eq:RootsCharPolynomial}. The perturbative part $ \mathcal P_1(z) $ gives a correction to  $z_\mu^{(0)} $ in leading order of $g^2$ and  is found by applying the PPT  in Eq.~\eqref{eq:polynomialPerturbationTheory} introduced in Appendix~\ref{sec:polynomialPerturbationTheory}.
In doing so, we find that the two roots $ z_1= z_1^{(0)} = -i \epsilon_1$ and $z_2= z_1^{(0)} = -i \epsilon_2$ remain unchanged in leading order, while the third root obtains a finite real part such that $z_{3} = - i E_1  - \gamma(E_1)+i \mathcal O \left( g ^2 \right)+ \mathcal O \left( g ^4 \right)$ with
\begin{eqnarray}
\gamma(\omega)  &=&  g^2  \nu_C ( \omega), \label{eq:energyResolvedRelaxationRate} 
\end{eqnarray}
where the cavity LDOS in Eq.~\eqref{eq:CavityLocalDensityOfStates} can be explicitly written as
\begin{eqnarray}
\nu_C ( \omega) &=&\frac{1}{\pi} \frac{\kappa\sigma}{\left(  - \omega + E_C - \kappa  \left( - \omega + E_M  \right)  \right)^2 + \kappa^2 \sigma ^2  } \nonumber
\label{eq:energyResolvedRelaxationRateLorentz}, \\
\kappa &=&  \frac{   g^2N  } { (- \omega + E_M) ^2 +  \sigma^2   } .
\end{eqnarray}
We note that Eq.~\eqref{eq:energyResolvedRelaxationRate}  agrees with the result in Ref.~\cite{Dubail2021}.
The imaginary  part shift $\propto g^2$ (i.e., the energy shift) can be neglected as it is not the leading order term.
 Using the roots to perform the inverse Laplace transformation, the Green's function becomes as a function of time 
\begin{eqnarray}
G_{1,1} (t )  &=&   A_1 e^{ z_1 t } + A_2 e^{ z_2 t }  + A_3 e^{ z_3 t },  \nonumber \\
A_{1}  &=&  \frac{   g^2  (z_1+ i E_M + \sigma )}{\left( z_{1} + i E_1 \right)^2 (z_{1} -z_{2}  ) } \rightarrow 0,     \nonumber\\
A_{2}  &=&  \frac{   g^2  (z_2+ i E_M \nonumber + \sigma )}{\left( z_{2} + i E_1 \right)^2 (z_{2} -z_{1}  ) }  \rightarrow 0 ,  \\
A_{3}  &=&  \frac{   g^2  (- i E_1 + i E_M + \sigma )}{\gamma(E) (- i E_1  -z_{1}  )  (- i E_1  -z_{2}  )  } \rightarrow 1  .
\end{eqnarray}
In the thermodynamic limit $N\rightarrow \infty , g\rightarrow 0$, which we have assumed throughout the calculation, the amplitudes $A_{1},A_{2}\rightarrow 0$, such that  the occupation of $n_1(t)$ is solely determined by the  real part of the root $z_3$. 
For consistency, we finally average $\gamma(E_1)$ in Eq.~\eqref{eq:energyResolvedRelaxationRateLorentz}  over the donor energy $E_1$, which is distributed according to the Lorentz distribution in Eq.~\eqref{eq:LorentzDistribution}, such that the disorder-averaged relaxation rate reads as
\begin{eqnarray}
\overline\gamma &=& \int dE_1  \gamma(E_1) P(E_1) \nonumber  \\
&=& \frac{g^2 }{\pi}  \frac{\sigma +  \frac{g^2N}{2\sigma } }{\left( E_M -E_C \right)^2 + \left( \sigma +  \frac{g^2N}{2\sigma }\right)^2 } ,
\label{eq:disorderAveragedRelaxationRate} 
\end{eqnarray}
whose physical behavior will be discussed in the next section.

 We note that the relaxation rate in Eq.~\eqref{eq:energyResolvedRelaxationRate}  is exact in the thermodynamic limit $g\rightarrow 0,N\rightarrow \infty$ and holds also for general disorder distributions. The derivation   is still valid for a Green's function with an arbitrary number of  poles. Possible branch cuts of the Green's function vanish in the thermodynamic limit as the second term in Eq.~\eqref{eq:disorderdAveragedGreensFunctionRelaxation} is proportional to $g^2$.

\subsection{Discussion of the relaxation dynamics}

\label{sec:discussionRelaxationRate}

\begin{figure}
	\includegraphics[width=1\linewidth]{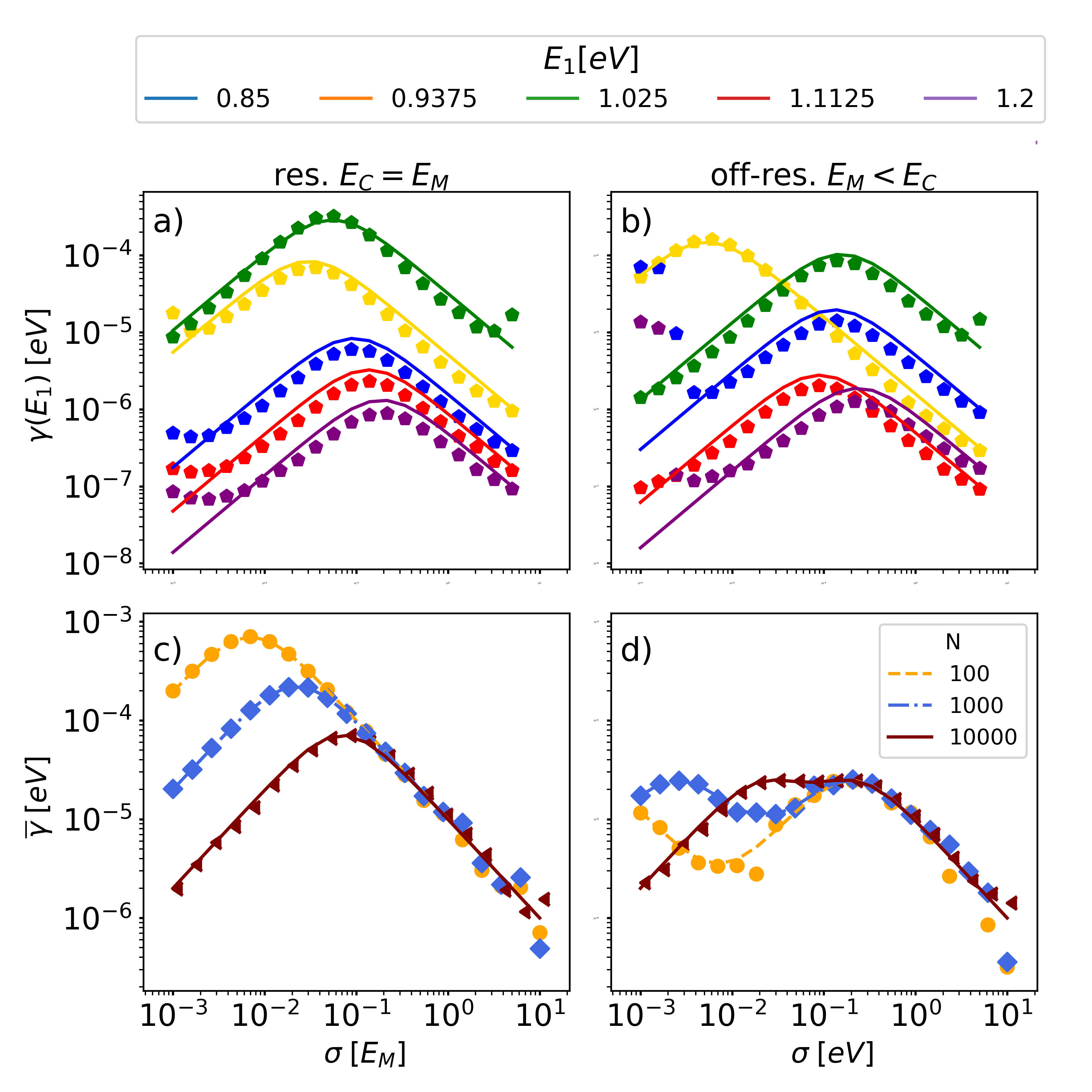}
	\caption{  (a,b) Energy-resolved relaxation rate $\gamma (E_1)$  for  the resonant system  $E_C= E_M =1\;\text{eV}$ (a), and  for the off-resonant system $E_C=1.05\;\text{eV} , E_M =0.95\;\text{eV}$ (b). In both (a) and (b), $g=0.001\;\text{eV}$ and $N=2000$. (c), (d)  Disorder-averaged relaxation rate $\overline \gamma$ for the same parameters as in (a) and (b) for different molecule numbers. Symbols represent the finite-size simulation to verify the analytical treatment, where the number of disorder samples $M_S$ is such that $M_S\cdot N  =10^6$.  }
	\label{figRelaxationRateDisorder}
\end{figure}

\begin{figure}
	\includegraphics[width=1\linewidth]{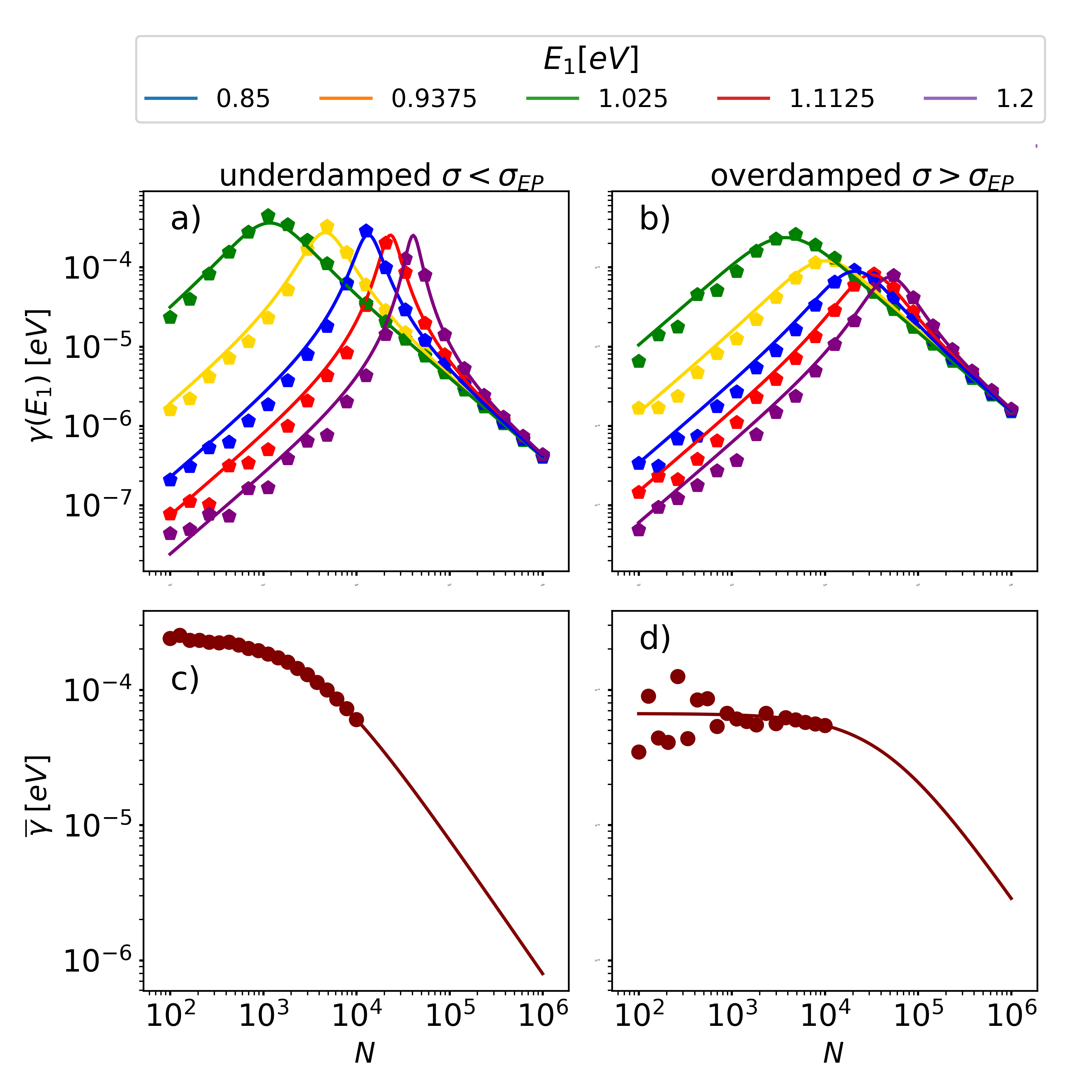}
	\caption{ Same as in Fig.~\ref{figRelaxationRateDisorder} but as a function for molecule number $N$ . The disorder in (a) and (c) is $\sigma =0.04\;\text{eV}$ and in (b) and (d) $\sigma =0.15\;\text{eV}$.  }
	\label{figRelaxationRateParticleNmb}
\end{figure}

The energy-resolved relaxation rate  in Eq.~\eqref{eq:energyResolvedRelaxationRateLorentz}   and the disorder-averaged relaxation rate in  Eq.~\eqref{eq:disorderAveragedRelaxationRate} are depicted  in Figs.~\ref{figRelaxationRateDisorder} and~\ref{figRelaxationRateParticleNmb}.  The solid lines depict the analytical results, while the symbols depict the numerical results of the  finite-size simulations, which are described in details in Appendix~\ref{sec:detailsNumericalEvaluationRates}. Several points are worthwhile to be discussed:

\textbf{Disorder dependence.} The energy-resolved relaxation rate $\gamma(E_1)$  is shown in  Fig.~\ref{figRelaxationRateDisorder}(a,b) as a function of disorder $\sigma$ for different values of $E_1$   for the resonant  $E_M = E_C $ and off-resonant  $E_M < E_C$ systems. 
We observe a turnover as a function of $\sigma$ for all values of $E_1$. For $E_1\neq E_M$ and/or $ E_M \neq E_C$, we find from Eq.~\eqref{eq:energyResolvedRelaxationRateLorentz} in the limiting cases 
 \begin{equation}
 \gamma(E_1)\propto  
 \begin{cases}
 \frac{\sigma}{N} & \sigma\ll g\sqrt{N} \\
 \frac N  \sigma  &  \sigma\gg g\sqrt{N}
 \end{cases}.
 \label{eq:energyResolvedRelaxationScaling}
 \end{equation}
 Only in the special case $E_1 =  E_M =E_C$,  $\gamma(E_1)$ is monotonically decreasing as a function of $\sigma$ (not shown). 
As $\gamma(E_1) \propto \nu_C(E_1) $, the relaxation rate depends  on the distance from the peak $\gamma(E_1) \propto \left(E_1 - \text{Re} \;\epsilon_\mu \right)^2$ and the  peak width  $\gamma(E_1) \propto \text{Im}\; \epsilon_\mu $ of the peaks in Figs.~\ref{figEigenStateAnalysis}(a),~\ref{figEigenStateAnalysis}(d),~\ref{figEigenStateAnalysis}(g), and~\ref{figEigenStateAnalysis}(j).

For example, in the resonant case $E_M=E_C$ for an energy $E_1 > E_M$ and $E_1 - E_M \ll\Omega_R$, the right peak in Figs.~\ref{figEigenStateAnalysis}(a) and~\ref{figEigenStateAnalysis}(d) related to the eigenenergy $\epsilon_2$ determines  $\gamma(E_1) \propto \nu_C(E_1) $.
 For small $\sigma$, we find from Eq.~\eqref{eq:asymptoticBehaviorEnergies} that $\text{Re} \;\epsilon_\mu -E_M \propto \sqrt{N}$ and $\text{Im}\; \epsilon_\mu \propto \sigma$, which explains the  behavior for small $\sigma $ in Eq.~\eqref{eq:energyResolvedRelaxationScaling}. Likewise for large $\sigma$, we find from Eq.~\eqref{eq:asymptoticBehaviorEnergies} that $\text{Re} \; \epsilon_\mu =  E_M = \text{const} $ and $\text{Im}\; \epsilon_\mu \propto N/\sigma$, which explains the  behavior for large $\sigma $ in Eq.~\eqref{eq:energyResolvedRelaxationScaling}.

\textbf{Energy dependence.} Overall, the relaxation rate is larger for energies $\left| E_1-E_C\right|\approx 0$. For the selected energies in Figs.~\ref{figRelaxationRateDisorder}(a) and~\ref{figRelaxationRateDisorder}(b), the energies $E_1= 0.9375\;\text{eV}$ and $E_1 = 1.025\;\text{eV}$ exhibit the largest relaxation rates, while the  energies $E_1= 0.85\;\text{eV}$ and $E_1 = 1.2\;\text{eV}$ exhibit the smallest relaxation rates. This is due to the shape of $\nu_C(E_1)$, which decays with  $1/\left(E_1 -E_C \right)^2$  away from the peaks according to Eq.~\eqref{eq:energyResolvedRelaxationRate}. We further observe that the maximum position in Figs.~\ref{figRelaxationRateDisorder}(a) and~\ref{figRelaxationRateDisorder}(b) as a function of $\sigma$ depends  on the value of $E_1$. For the resonant system, the maxima are located around the same value, while  for the off-resonant system, the maximum for the $E_1=0.9375\;\text{eV}$ curve is reached earlier than the other curves.

\textbf{Disorder-averaged relaxation rate.}
The disorder-averaged relaxation rate  is depicted   in Figs.~\ref{figRelaxationRateDisorder}(c) and~\ref{figRelaxationRateDisorder}(d)   as a function of  $\sigma$  for different molecule numbers. In Fig.~\ref{figRelaxationRateDisorder}(c) for the resonant case, we observe a turnover which is a consequence of the turnovers of the energy-resolved relaxation rates in Fig.~\ref{figRelaxationRateDisorder}(a).
According to the exact expression in Eq.~\eqref{eq:disorderAveragedRelaxationRate}, the disorder-averaged relaxation rate  scales as
\begin{eqnarray}
\overline \gamma \propto
\begin{cases}
\frac{\sigma} {N} & \sigma\ll g \sqrt{N} \\
 \frac{1 } {\sigma} & \sigma\gg g \sqrt{N}
\end{cases},
\label{eq:currentScaling}
\end{eqnarray}
in the two limiting cases, which  is similar to Eq.~\eqref{eq:energyResolvedRelaxationScaling} except for a factor $N$ for large disorder and can be clearly recognized in Fig.~\ref{figRelaxationRateDisorder}(c). 
For small $\sigma$, the width of the Lorentzian distribution  increases with $\sigma$, which results in an increasing overlap with   the cavity LDOS, whose spectral width  also increases according to the imaginary part  of the eigenvalues in Fig.~\ref{figRootsNoDephasing}.  For large $\sigma$, the cavity LDOS $\nu_C ( E_1) $ becomes very narrow as discussed in Sec.~\ref{sec:CavityLocalDensityOfStates} and shown in Fig.~\ref{figEigenStateAnalysis}(d), while the Lorentz distribution broadens and scales as $P(E_1) \propto 1/\sigma  $ for $E_1 \approx E_M$. As a defining property, the integral of $\nu_C(E_1)$ over all energies equals  one, such that the overall relaxation rate scales as in Eq.~\eqref{eq:currentScaling}.

The off-resonant case depicted in Fig.~\ref{figRelaxationRateDisorder}(d) exhibits two maxima for $N=100$ and $N=1000$. This is a consequence of the energy-resolved relaxation rate, whose peak position depends on the donor energy as can be seen in Fig.~\ref{figRelaxationRateDisorder}(b). Overall, the exact shape of the disorder-averaged dissipation rate sensitively depends on the shape of $\nu_{C}(E_1)$ and thus on the system parameters.

\textbf{Molecule-number dependence.} The energy-resolved relaxation rate is depicted in Figs.~\ref{figRelaxationRateParticleNmb}(a) and~\ref{figRelaxationRateParticleNmb}(b) as a function of  molecule number in the underdamped and overdamped regimes for the resonant system $E_M =E_C$. In both panels we observe a similar behavior. Interestingly, the rates exhibit a turnover in both regimes. The maximum position sensitively depends on the donor energy $E_1$, where the  maximum for small $\left|E_1 -E_C\right| $ is reached earlier than for large $\left|E_1 -E_C\right| $. The limiting cases of $\gamma(E_j)\propto N$ for small $N$ and $\gamma(E_j)\propto 1/N$ for large $N$ can be directly inferred from Eq.~\eqref{eq:energyResolvedRelaxationRateLorentz}. For large $N$, the relaxation rate  $\gamma(E_1)$ becomes independent off $E_1$ as this  limit is equivalent to a rescaling of the system energies $ E_1/ g\sqrt{N} \rightarrow 0$.

The disorder-averaged relaxation rate $\overline{\gamma}$ is depicted in Figs.~\ref{figRelaxationRateParticleNmb}(c) and~\ref{figRelaxationRateParticleNmb}(d). For small $N$, the rate $\overline{\gamma}$ is approximately constant in agreement with Eq.~\eqref{eq:currentScaling}. 
For large  $N $ (i.e., $g\sqrt{N} \gg \sigma$), the  factor $1/N$   can be explained by the shape of $\nu_C(\omega)$  discussed in Sec.~\ref{sec:CavityLocalDensityOfStates}: for the resonant system, the polariton peaks in Fig.~\ref{figEigenStateAnalysis}(a) are located around $E_{peak} = E_M \pm g\sqrt{N}$ and decay as $\nu_C(E)\propto 1/(E - E_{peak} )^2 $ with increasing distance from the peak center. When the donor energy is  distributed around $E_M$ for small $\sigma/ g\sqrt{N}$, then the cavity mode overlap contributes the extra factor $1/N$.

\textbf{Finite-size simulation.}
In  Figs.~\ref{figRelaxationRateDisorder} and \ref{figRelaxationRateParticleNmb}, the analytical solutions are compared with finite-size simulations, which are represented as symbols. Overall, both calculations agree very well with each other, except for deviations for very small $\sigma$ in Figs.~\ref{figRelaxationRateDisorder}(a) and~\ref{figRelaxationRateDisorder}(b). The deviations occur for donor energies $E_1$ away from the center of the Lorenz distribution $E_M$. For these energies, the density of states is very low, such that the continuum limit, which has been applied in the derivation, is not justified. This problem is even more prominent when the donor energy $E_1$ is close to $E_C$ as can be seen for $E_1 = 1.125\;\text{eV}$ and $2.0\;\text{eV}$ in Figs.~\ref{figRelaxationRateDisorder}(b), as the level repulsion leads to a depletion of the cavity LDOS in this region. However, these deviations are  not visible in the disorder-averaged rate as the energies in the sparse spectral regime  hardly contribute to the disorder average. In Fig.~\ref{figRelaxationRateParticleNmb} we clearly observe the improvement of the analytical calculation  when approaching the thermodynamic limit $N\rightarrow \infty$.

\section{Transport  }
\label{sec:excitationTransfer}

Here, we consider the  transport of  an excitation which is  initially located at the donor $j=1$  to  the reservoir which is coupled to the acceptor molecule $j=N$. The  transport process is sketched in Fig.~\ref{figSketchCavity}(c) and consists of three phases: excitation of the donor, relaxation as considered in Sec.~\ref{sec:relaxtionDynamics}, and trapping at the reservoir. The first two steps are fast compared to the trapping process, as the excitation has to \textit{find} the acceptor  in a random-walk fashion, i.e., it scales with number of molecules $N$. The  transport rate, which we consider in the following, is defined as the inverse of the mean first-passage time of the trapping process.

The derivation involves the following two steps: In Sec.~\ref{sec:transportHamiltonian}, we introduce the  transport Hamiltonian, which is amended by the acceptor reservoir, and derive an effective characteristic polynomial describing the  transport dynamics.   In Sec.~\ref{sec:rootsCharacteristicPolynomial}, we determine the roots of the effective characteristic polynomial, whose real parts determine the   transport rate. For the analytic treatment, we apply the PPT and the ESM introduced in Appendix~\ref{sec:AnalyticalTechniquesDetails}.
 Readers only interested in the physical content can skip the derivations and directly proceed to the discussion in Sec.~\ref{sec:DiscussionMeanFirstPassageTime}.

\subsection{Transport Hamiltonian and Green's function}
\label{sec:transportHamiltonian}

To calculate the  transport dynamics, we couple the Hamiltonian in Eq.~\eqref{eq:lightMatterHamiltonian} to an additional reservoir at the acceptor  $j=N$ as depicted in Fig.~\ref{figSketchCavity}(c). The amended Hamiltonian  is then given by
\begin{eqnarray}
H_{tr}&=& H + \hat H_{R} + \hat H_{AR}\nonumber, \\
\hat H_{R} &=& \sum_{k} \epsilon_{k }\hat c_{k}^\dagger \hat c_{k} ,   \nonumber  \\   
\hat H_{AR} &=& \sum_{k=1}^{N_R} g_R \hat B_{N}^\dagger \hat c_{k} + \text{H.c.},   
\label{eq:hamiltonianTranport}
\end{eqnarray}
where the reservoir is described by the operators $\hat c_{k}$, which are coupled with strengths $g_R$ to the acceptor.
The transport behavior of this and similar Hamiltonians has been numerically investigated in Refs.~\cite{Chavez2021,Feist2015,Schachenmayer2015,Wellnitz2021,Hagenmueller2017}. 

Initially, the system is excited at the donor  $j=1$ such that the relevant  Green's functions  are 
\begin{eqnarray}
	G_{C,1}(z)  &=&  -i  \frac{   g }{\left( z + i   E_C (z) \right) \left( z + iE_1 \right) } \label{eq:fullSolutionLaplaceSpaceOld}   \\
&+&i  \frac{  g^3   }{\left( z + i   E_C (z) \right)^2\left( z+ i  E_N(z) \right)   \left( z + iE_1 \right) } \nonumber ,  \\
G_{j\neq N,1}(z) &=& \frac{\delta_{j,1} }{z + i E_1} \nonumber \\
& -& \frac{ g^2    }{\left( z + i E_j \right)\left( z + i   E_C (z) \right) \left( z + iE_1 \right) } \nonumber  \\
 & &\hspace{-0.2\linewidth } + \frac{ g^4  }{\left( z + i E_j \right)\left( z + i   E_C (z) \right)^2\left( z+ i  E_N(z) \right)   \left( z + iE_1 \right) } \nonumber ,   \\
G_{ N,1}(z)  &=&   -  \frac{ g^2  }{\left( z + i   E_N (z) \right)\left( z + i   E_C (z) \right) \left( z + iE_1 \right) } ,\nonumber  
\end{eqnarray}
where the two  auxiliary functions are defined by
\begin{eqnarray}
E_C (z) &=&  E_C  +  \sum_{j=1}^{N-1} \frac{g^2}{z + i E_j  }  , \\
E_N (z) &=&  E_N  + \sum_k \frac{g_R^2}{z + i E_k }  +  \frac{g^2}{z + i  E_C (z) }. \label{eq:acceptorAuxilaryFunktion}  
\end{eqnarray}
For an appropriate multiplication with the factors $(z + iE_j) $ and $(z + iE_k)  $, the denominators of the respective last terms in Eq.~\eqref{eq:fullSolutionLaplaceSpaceOld} define the polynomial
\begin{eqnarray}
\mathcal P (z)&=&  \left( z+ i  E_N(z) \right)  \left(  z + i  E_C (z)  \right)\nonumber   \\
&\times&  \prod_j^{N} (z + iE_j)\prod_k^{N_R} (z + iE_k),
\label{eq:fullCharacteristicPolynomial}
\end{eqnarray}
which is equivalent to the characteristic polynomial of the Hamiltonian in Eq.~\eqref{eq:hamiltonianTranport} (with a replacement of $z\rightarrow - i E$). It is not possible to find all roots of this  polynomial, which has order $N+N_R+1$. As in  Sec.~\ref{sec:effectiveHamiltonian}, we can take the thermodynamic limit in the second term of Eq.~\eqref{eq:acceptorAuxilaryFunktion}. Under the assumption that  $E_k$ is distributed according to the Lorentz distribution, we can simply replace $E_k \rightarrow E_R +  \Sigma$, where $E_R$ denotes the center of the reservoir distribution and $\Sigma$ is its width. Yet, as the  transport rate is determined by the  microscopic dynamics of the molecular excitations, the disorder average  is not applied to the auxiliary function $E_C(z)$. After the disorder average of the reservoir states, the effective characteristic polynomial reads
\begin{eqnarray}
\mathcal P (z) &=&    \left(  z + i  E_C (z)  \right) \left( z+ i  E_N(z) \right)   (z + iE_R + \Sigma) \nonumber \\
&\times&\prod_j^{N-1} (z + iE_j) ,
\label{eq:effectiveCharacteristicPolynomial}
\end{eqnarray}
whose $N+2 $ complex roots $z_\mu$ determine the inverse Laplace transformation in Eq.~\eqref{eq:inverseLaplaceTranformation}.

\subsection{Transport rate}

\label{sec:rootsCharacteristicPolynomial}

To determine the  transport rate, it is not necessary to calculate the exact time evolution of the Green's functions in Eq.~\eqref{eq:fullSolutionLaplaceSpaceOld}, as the coherent  dynamics does not change the occupation within the system  $\hat H$. 
The roots of Eq.~\eqref{eq:effectiveCharacteristicPolynomial} determine the  transport rates $\Gamma_\mu =- \text{Re}\; z_\mu$ corresponding to the energies $E_\mu= -\text{Im}\; z_\mu$.
In the thermodynamic limit $g\rightarrow 0,N\rightarrow\infty$, the  transport rate $\Gamma(E)$ is  a smooth function  (after an appropriate disorder average over the eigenstates in an infinitesimal energy interval $\left[E,E+dE \right]$). Because of the weak coupling $g$,  the donor state is a superposition of  eigenstates  with $E \approx E_1$, such that the energy-resolved  transport rate is given by $\Gamma(E_1)$.

The roots $z_\mu$ of the characteristic polynomial  in Eq.~\eqref{eq:effectiveCharacteristicPolynomial} cannot be determined analytically  as the order of the polynomial is $N+2$. However, we can evaluate the roots in a stochastic fashion, which is sufficient to calculate  the  transport rate in the thermodynamic limit. To this end, we apply the ESM and the PPT introduced in Appendix~\ref{sec:AnalyticalTechniquesDetails}:
we partition the effective   characteristic polynomial  in Eq.~\eqref{eq:effectiveCharacteristicPolynomial} as 
\begin{eqnarray}
\mathcal P (z ) &=& \mathcal P_0 (z ) + \mathcal P_1 (z ) \\
\mathcal P_0 (z ) &=& \prod_{\mu =1}^{N} \left( z + i E_\mu \right) \nonumber,  \\
&\times& \left[ \left( z + i E_N (z)\right) \left( z  + i E_R  +   \Sigma \right)  + N_R g_R^2  \right]   , \label{eq:effectivCharP0}  \nonumber \\
\mathcal P_1 (z ) &=& \left[g^2 \sum_\mu \frac{s_\mu  }{z + i E_\mu } \right]    \prod_\mu  \left( z + i E_\mu \right) \left( z  + i E_R  +   \Sigma \right), \label{eq:effectivCharP1}\nonumber
\end{eqnarray}
which defines  the unperturbed and perturbation parts of the characteristic polynomial. The term 
\begin{eqnarray}
\prod_{\mu =1}^{N}&&\left( z + i E_\mu \right) \\
&=& \left( z + i E_C  +  \sum_{j=1}^{N-1} \frac{g^2}{z + i E_j  } \right)  \prod_{j =1}^{N-1}\left( z + i E_j \right)\nonumber
\end{eqnarray}
 represents the factorized characteristic polynomial of the molecules $j=1,...,N-1$ and the cavity mode under the assumption that $g_N\rightarrow 0$. This is  a formal representation, where the corresponding roots $-iE_\mu$ are assumed to be given. The second factor of $ \mathcal P_0 (z )$ represents the acceptor $j=N$ and its coupling to the reservoir.  The unperturbed roots of $\mathcal P_0 (z )$ are thus given by
\begin{eqnarray}
z_{\mu}^{(0)} &=& - i E_\mu \nonumber ,  \\
z_{(N+1),(N+2)}^{(0)} &=&-\frac{1}{2}\left(iE_A +iE_R^{(\Sigma)}  \right) \nonumber \\
&\pm &   \frac{1}{2}\sqrt{\left(iE_A - iE_R^{(\Sigma)} \right)^ 2 - 4 g_R^2 N_R  } \nonumber, \\
\end{eqnarray}
where $E_R^{(\Sigma)} = E_R - i \Sigma$. The perturbative polynomial term $\mathcal P_1 (z )$ describes the coupling between the cavity system and the acceptor. To enable an analytical treatment we have applied the  ESM to the term 
\begin{eqnarray}
\frac{1}{z + i E_C (z) }  \rightarrow  \sum_\mu \frac{s_\mu  }{z + i E_\mu }    ,  
\label{eq:esrApplicationGreensFkt}
\end{eqnarray}
appearing in the auxiliary function in Eq.~\eqref{eq:acceptorAuxilaryFunktion}.
After identifying the left-hand side of Eq.~\eqref{eq:esrApplicationGreensFkt} with $G_{C,C}(z)$  in Eq.~\eqref{eq:greensFktDD}, comparing with the definition of the LDOS in Eq.~\eqref{eq:localDensityOfStates}, and applying the ESM  rule in Eq.~\eqref{eq:expansionCoeffients}, we find that the expansion coefficients are $ s_\mu = \nu_C(E_\mu)/\nu(E)$. In  Eq.~\eqref{eq:esrApplicationGreensFkt} we have used the same $E_\mu$ as in the formal expression of $\mathcal P_0 (z )$.

Using the PPT given in Eq.~\eqref{eq:polynomialPerturbationTheory}  of Appendix~\ref{sec:polynomialPerturbationTheory}, we can find the corrections $\delta_\mu$ to the roots $z_\mu^{(0)}$ resulting from the perturbative polynomial.  In the lowest order of $g$ the roots of $\mathcal P (z )$ then read as $z_\mu = z_\mu^{(0)} +\delta_\mu$, where
\begin{eqnarray}
\delta  _{\mu} 
&=& g ^2 s_\mu  \frac{1 }{  - i E_\mu + i E_N   +  \frac{ N_R G_R^2}{ - i E_\mu  + i E_R +\Sigma }  } + \mathcal O \left( g ^4 \right).\nonumber \\
\end{eqnarray}
Replacing  $s_\mu$  and  identifying the fraction as the local density of states of the acceptor  $\nu_N(E)$, we obtain the  roots in the thermodynamic limit
\begin{eqnarray}
z_{\mu} &=& - i E_\mu  - \Gamma(E_\mu) + i\mathcal O \left( g^2 \right) +  \mathcal O \left( g^4 \right) , \nonumber \\
z_{(N+1),(N+2)} &=& z_{(N+1),(N+2)}^{(0)}+ i\mathcal O \left( g^2 \right) + \mathcal O \left( g^2 \right),
\label{eq:rootsThermodynamicLimit}
\end{eqnarray}
where 
\begin{eqnarray}
\Gamma(E) &=&   g^2 \frac{\nu_C (E)}{\nu(E) }     \nu_N (E) 
\label{eq: transportRate} 
\end{eqnarray}
is the  transport rate. We have neglected the correction in the imaginary part in $z_{\mu}$ and the complete shift in $z_{(N+1),(N+2)}$ as they have a vanishing influence in the thermodynamic limit $g\rightarrow 0$. Yet, the finite real part $\Gamma(E) $ describes the exponential decay of the system occupation and thus determines the  transport rate.  We note that, even though  a Lorentz distribution of the reservoir state energies $E_k$ is assumed, the derivation can be straightforwardly generalized to more general disorder distributions of the reservoir, for which additional roots and possible branch cuts have no influence on the final outcome.

\subsection{Discussion of the  transport rate}
\label{sec:DiscussionMeanFirstPassageTime}

\begin{figure}
	\includegraphics[width=1\linewidth]{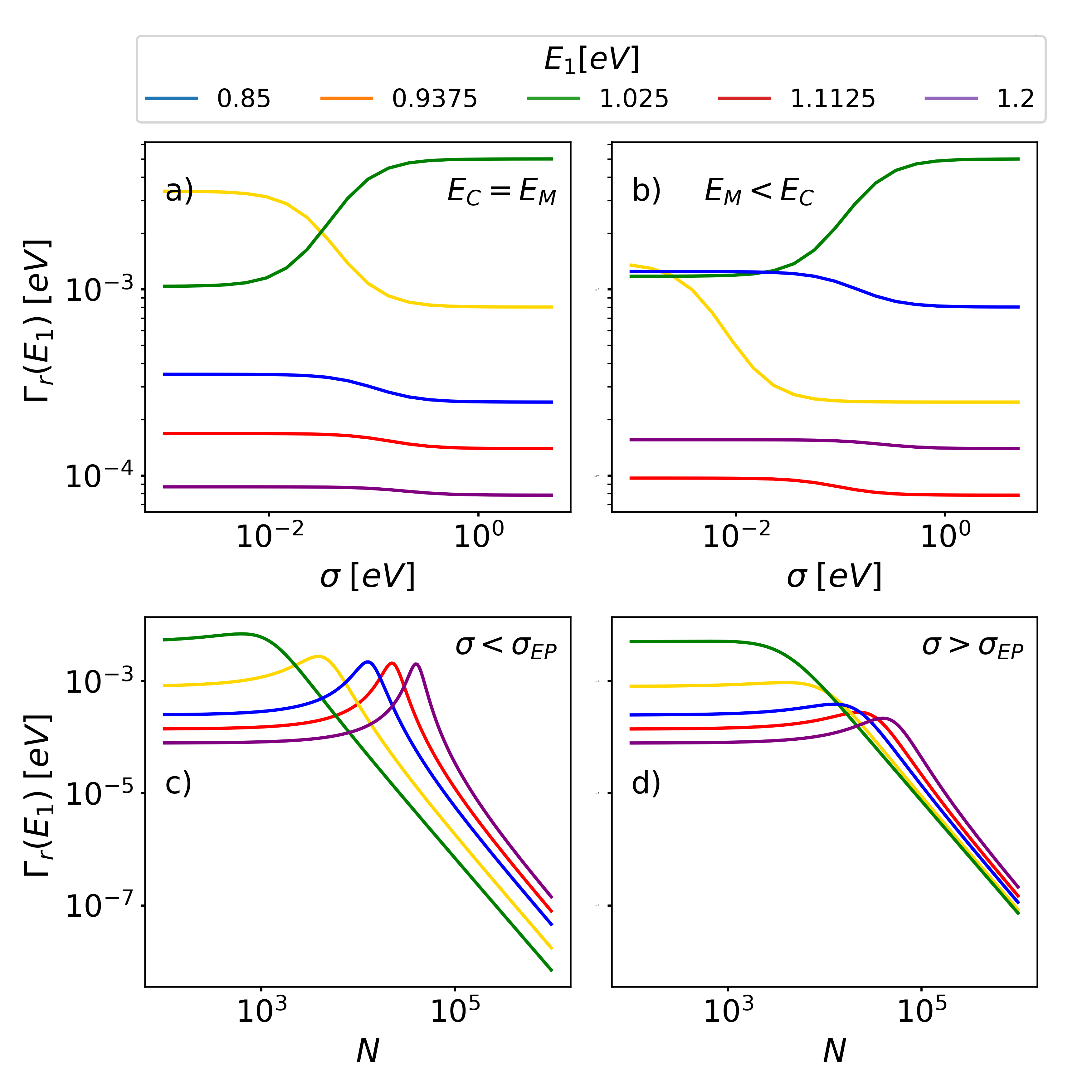}
	\caption{ Resonant   transport rate in Eq.~\eqref{eq:resonantEnergyTransferRate} as a function of disorder (a), (b) and molecule number (c), (d). The light-matter coupling is $g=0.001\;\text{eV}$ in all panels. The specific parameters are  (a) $E_C=E_M=1\;\text{eV}, N=2000$, (b) $E_C=1.05\;\text{eV}, E_M=0.95\;\text{eV}, N=2000$, (c) $E_C=E_M=1\;\text{eV}, \sigma=0.04\;\text{eV}$, (d) $E_C=1.05\;\text{eV}, E_M=0.95\;\text{eV}, \sigma=0.15\;\text{eV}$. The disorder-averaged  transport rate $\overline \Gamma$ has the same functional dependence as the disorder-averaged relaxation rate $\overline \gamma$ in Figs.~\ref{figRelaxationRateDisorder} and \ref{figRelaxationRateParticleNmb}, i.e, $\overline \Gamma =\overline \gamma/N$. }
	\label{figTransferRate}
\end{figure}

We emphasize that  the  transport rate  in Eq.~\eqref{eq: transportRate}  is exact in the thermodynamic limit. We recall that the  transport rate has to be evaluated at the donor energy, i.e.,  $\Gamma(E_1)$.
Its inverse   $\tau(E_1)= 1/\Gamma(E_1)$ denotes the time to deplete the corresponding eigenstates, i.e., the mean first-passage time. Recalling that $\nu_N(E)$ and $\nu_C(E)$ are proportional to the absorption of the acceptor and the cavity,  the transport rate thus resembles   the celebrated F\"orster rate, weighted by  the inverse density of states $\nu (E)$. The following points are worthwhile to discuss in details.

\textbf{Resonant   transport rate.} 
For a weak reservoir coupling, the acceptor LDOS is strongly peaked around $E_N$. For the following discussion, we consider a resonant donor-acceptor configuration with $E_1 = E_N$ and assume  the acceptor LDOS to be  a constant $ \nu_N (E_1) =\nu_0 $ for simplicity. The  resulting resonant energy  transport rate 
\begin{equation}
\Gamma_r(E_1) =   g^2\nu_0  \frac{\nu_C (E_1)}{\nu(E_1) } 
\label{eq:resonantEnergyTransferRate}
\end{equation}  
  is proportional to the relaxation rate in Eq.~\eqref{eq:energyResolvedRelaxationRate}, but renormalized with the total density of states. The renormalization describes a competition of the eigenstates for the overlap with the cavity mode. The more eigenstates exist at a given energy, the smaller is the overlap with the cavity mode, which suppresses the coherent coupling between donor and acceptor.

\textbf{Disorder dependence of the resonant   transport rate.} Figures~\ref{figTransferRate}(a) and~\ref{figTransferRate}(b) plot $\Gamma_r(E_1) $ as a function of the disorder in the resonant $E_M=E_C$ and off-resonant $E_M<E_C$ systems. In both cases, the overall behavior is qualitatively the same.
We observe that   $\Gamma_r(E_1) $ is  independent off the disorder in the small- and large- disorder regimes. The limits of the total density of states (which converges to a Lorentz function in the thermodynamic limit, cf. Eq.~\eqref{eq:systemDensityOfStates}) are $\nu(E_1\neq E_M) \propto \sigma $ for small $\sigma$ and $\nu(E_1\neq E_M) \approx 1/\sigma $ for large $\sigma$. Combining this with the limiting cases of $\nu_C(E_1)$ given in Eq.~\eqref{eq:energyResolvedRelaxationScaling} explains the disorder-independent regimes. Thus,   the  overlap  of the individual eigenstates with the cavity remains constant in the small- (large-) disorder regime due to the cancellation of the simultaneous increase (decrease) of the density of states and the increase (decrease) of the cavity LDOS.
  
\textbf{Molecule-number  dependence of the resonant   transport rate.}  Figures~\ref{figTransferRate}(c) and~\ref{figTransferRate}(d) plot $\Gamma_r(E_1) $ as a function of the molecule number in the underdamped  and overdamped regimes. In both  regimes, the resonant  transport rate exhibits a similar behavior. For small molecule numbers,  $\Gamma_r(E_1)$ is  independent off $N$, while it decreases as $N^{-2}$ for large molecule numbers. Combining the scaling properties of the density of states $\nu(E)\propto N$ with the limiting cases in Eq.~\eqref{eq:energyResolvedRelaxationScaling}, explains these observations. Thus, for small molecule numbers the scalings of the cavity LDOS and the total density of states cancel, while for large molecule numbers  the scalings contribute constructively.

\textbf{Disorder-averaged  transport rate.}  For a weak reservoir coupling, the acceptor LDOS is strongly peaked such that it converges to $\nu_N(E) \rightarrow \delta(E-E_n)$ in the thermodynamic limit.  Assuming that $E_1$ and $E_N$  are  distributed according to the Lorentz distribution in Eq.~\eqref{eq:LorentzDistribution}, we can average the energy-resolved  transport  rate in Eq.~\eqref{eq: transportRate}  to obtain the disorder-averaged  transport rate 
\begin{eqnarray}
  \overline \Gamma &=& \int dE_1 \int dE_N \Gamma(E_1) P(E_N) P(E_1) \nonumber \\
  &=& \frac{1}{N } \int dE_1  \nu_C(E_1)  P(E_1) \nonumber \\
  &=& \frac{1}{N } \overline \gamma ,
  \label{eq:disorderAveragedTransferRate}
\end{eqnarray}
which is thus direct proportional to the disorder-averaged relaxation rate in Eq.~\eqref{eq:disorderAveragedRelaxationRate}. We note that the disorder-averaged transport rate in Eq.~\eqref{eq:disorderAveragedTransferRate} considers independent  donor and acceptor energies, which is different from  the resonant  transport rate in Eq.~\eqref{eq:resonantEnergyTransferRate}.
 Formally, the factor $1/N$ appears because of the factor $\nu(E)\rightarrow N P(E)$ in Eq.~\eqref{eq: transportRate}.
The inverse of the rates  $\tau_\gamma = \frac 1 {\overline \gamma} $ and $\tau_\Gamma = \frac 1 {\overline \Gamma} $ are  the mean first-passage times of the relaxation and  transport processes.  In a classical picture, a jump from one molecule to another takes $\tau_\gamma$. In a random walk fashion, the number of jumps to reach the acceptor  is on the order of $N$, so the mean  first-passage time is $\tau_\Gamma  \propto N\tau_\gamma$.  The quantum calculation arrives at the exact relation  $\tau_\Gamma  = N\tau_\gamma$. We note that the derivation in Sec.~\ref{sec:rootsCharacteristicPolynomial} can be straightforwardly generalized to a system with $N_A$ acceptors as long as $N_A \ll N$. In doing so, one finds that $\overline \Gamma =  N_A / N \overline \gamma$, which underpins the interpretation of the  transport picture as a quantum random walk process.

Because of the close relation of the disorder-averaged relaxation and transfer rates in Eq.~\eqref{eq:disorderAveragedTransferRate}, the qualitative discussion in Sec.~\ref{sec:discussionRelaxationRate} is valid also for the transport process. The turnover as a function of disorder in Fig.~\ref{figRelaxationRateDisorder}(c) appears to be reminiscent of the turnover as a function of system-environment coupling or temperature, which is often associated with environment-assisted transport~\cite{Kassal2013,Ishizaki2012,Lambert2013,Scholes2014,Rebentrost2009,Chuang2016,Lee2015,Moix2013a}. However, we emphasize that the underlying physical mechanisms are different.      For weak disorder, the cavity-mediated transport observed in Fig.~\ref{figRelaxationRateDisorder}(c) is enhanced because of an increasing overlap of the cavity local density of states and the quantum emitter energy distribution. For large disorder, the cavity-mediated transport vanishes as the quantum emitter energy is increasingly distributed over a larger energy region, such that fewer quantum emitters are energetically resonant with the cavity mode. Consequently, quantum emitters and cavity mode decouple such that the cavity mode is subject to less decoherence.
%
%

\section{Discussion and  outlook}

\label{sec:discussion}

\subsection{Summary}

Using the Green's function method to  analytically solve the Fano-Anderson model, we have predicted the spectroscopic, relaxation and  transport features of polaritons in microcavities in the presence of energetic disorder. The central physical findings of this article are summarized as follows:
\begin{itemize}
	\item \textbf{Complex eigenenergy.} The average over energy disorder results in an effective Hamiltonian, which exhibits an exceptional point in its eigen solutions in the resonance case (i.e. $E_C=E_M$). This exceptional point defines two dynamical regimes: underdamped coherent dynamics in the weak disorder regime, where the decay rate increases linearly with disorder and the collective Rabi frequency decreases quadratically with disorder, and overdamped bi-exponential dynamics in the strong disorder regime, where the slow decay rate decreases with disorder and the fast decay rate increase with disorder.
	\item \textbf{Spectroscopy.} The contributions of the cavity mode and the bright state to the eigenstates of the disordered system define the cavity LDOS, $\nu_{C}$, and bright-state LDOS, $\nu_{BS}$,  which can be measured via the cavity absorption and the matter absorption, respectively.  In the weak disorder regime, the complex eigen solutions leads to two spectral peaks separated by the effective Rabi splitting. In the strong disorder regime, the cavity spectrum exhibits a central peak dictated by the slow eigen solution, whereas the matter spectrum results from the destructive interference of the two eigen solutions.
	%
	%
	 Intriguingly, the matter spectrum exhibits a complete absorption suppression at the energy of the cavity mode which is reminiscent of the  electromagnetically-induced transparency~\cite{Fleischhauer2005,Engelhardt2021} and the related vacuum-induced transparency effects~\cite{Field1993,Litinskaya2019}. In contrast to these effects, which appear in individual  atoms or molecules, the absorption suppression observed in $\nu_{BS}$ of Fig.~\ref{figEigenStateAnalysis} is a consequence of the collective destructive interference of the two-level quantum emitters and the cavity mode.
	\item \textbf{Energy-resolved relaxation rate.} For all donor energies $E_1 $, the energy-resolved relaxation rate, $\gamma(E_1) $,  exhibits a turnover as a function of disorder or molecule number.  Specifically, we have $\gamma(E_1) \propto \sigma/N $  in the weak disorder regime of $\sigma \ll g\sqrt{N} $ and $\gamma (E_1)  \propto N/\sigma $  in the strong disorder regime of  $\sigma \gg g\sqrt{N} $.

	\item \textbf{Disorder-averaged relaxation rate.} The turnover in $\gamma(E_1)$ translates into a turnover in the disorder-averaged relaxation rate 
	$\overline{\gamma} $ as a function of disorder but a monotonic decay of $\overline{\gamma}$ as a function of the molecule number.  Further, the disorder average modifies the scaling behavior.  For $\sigma\ll g\sqrt{N}$ we find $\overline \gamma \propto \sigma/N$, while for  $\sigma\gg g\sqrt{N}$ we find  $\overline \gamma \propto  1 / \sigma$.
	\item \textbf{Resonant   transport rate.} For the  transport from a donor to an acceptor which are in resonance $E_1 = E_N$, the rate is proportional to $\Gamma_r(E_1)\propto \nu_C(E_1)/\nu (E_1)$. Because of the presence of the total density of states $\nu (E_1)$, for $\sigma\ll g \sqrt{N}$,  the resonance transport rate decreases with the molecule number as $\Gamma_r(E_1)\propto 1 / N^2$, whereas for  $\sigma \gg g \sqrt{N}$, $\Gamma_r(E_1)$ is independent of both disorder and molecule number.   
	\item \textbf{Disorder-averaged transport rate.} 
	The disorder-averaged relaxation rate $\overline \gamma$ and transport rate $\overline \Gamma$ are related as $\overline \Gamma = \overline \gamma/N $. 
	The relaxation process is from the donor to molecule ensemble, whereas the  transport process is from donor to acceptor, one of the $N$ molecules, which explains the factor $N$. Overall, the relaxation and transport depend quadratically on the light-matter interaction $g$ when keeping the Rabi frequency $\Omega = 2g \sqrt{N}$ constant. 
\end{itemize}

\subsection{Discussion}

In the following, we elaborate on the  methods and  physical pictures established in this article. 

\textbf{Analytical methods.} 
The Green's functions approach is a flexible tool which yields rich insight into the polarition dynamics. The disorder average  enables  the derivation of an effective Hamiltonian,  which reduces a continuum of states to a non-Hermitian two-state Hamiltonian. Its complex-valued eigenenergies represent the damping dynamics of the two polaritons. This non-Hermitian feature accurately describes the mixing of the bright and dark states induced by the disorder and predicts the spectroscopic properties in the thermodynamic limit. The Green's function approach applied here allows for compact derivations of various observables on an equal footing, which clearly reveal the underlying physics.

The relaxation rate is calculated by evaluating the imaginary part of the  complex eigenenergies. To this end, we have developed the polynomial perturbation theory (PPT), which unifies the degenerate and  non-degenerate perturbation theories.
The calculation of the transport rate requires to evaluate $N+2$ roots of an effective characteristic polynomial. As this is analytically infeasible,  we  have developed  the exact stochastic mapping (ESM), which maps one sample of parameters to another sample of the same stochastic properties, but with a more convenient structure for the further  analytical treatment.

\textbf{Bright and dark states.} We have shown that the total density of states $\nu(E)$ can be written as a sum of the cavity LDOS,  $ \nu_{C}(\omega)$, the bright-state LDOS,  $ \nu_{BS}(\omega)$, and the dark-state LDOS, $ \nu_{DS}(\omega)$,   as explicitly given in Eq.~\eqref{eq:systemDensityOfStates}.  The bright-state and dark-state LDOS quantify the contributions of the bright and dark states of the homogeneous system to  the eigenstates of the disordered system. The spectral shape of  $ \nu_{C}(\omega)$  and $ \nu_{BS}(\omega)$ reflect the mixing of the bright state to the dark states in the presence of disorder.

The components of the total density of states can be accessed via spectroscopy. A realistic spectroscopic experiment simultaneously measures the cavity and matter absorption, i.e., $\chi(\omega) =\alpha_C \chi_C(\omega) + \alpha_M \chi_M(\omega)$ with  coefficients $\alpha_C $ and $\alpha_M$. One can use the scaling properties  $\chi_C(\omega)  \propto  \nu_C (\omega)  $ and  $ \chi_M(\omega)\cdot \propto V \nu_{BS} (\omega) $  to differentiate between the two contributions in a spectroscopic experiment by varying the volume $V$ while keeping the molecule density constant.

\textbf{Relaxation and  transport.}
The relaxation and  transport rates are primarily determined by the cavity LDOS. As the functional shape of the cavity LDOS is strongly influenced by the dark states, they have a substantial impact on the relaxation and transport properties.
The analytic treatment predicts a simple relation between the relaxation and  transport processes: $\Gamma_r(E_1) \propto  \gamma(E_1)/\nu(E_1) $, where $\Gamma_r(E_1)$ is the resonant  transport rate and $\gamma(E_1)$ is the energy-resolved relaxation rate.
We have analytically explained the behavior of the relaxation and  transport processes in the limiting cases of $\sigma \ll g\sqrt{N}$ and $\sigma \gg g\sqrt{N}$, respectively. Interestingly, the resonant   transport rate is independent of disorder and molecular number in the strong disorder regime $\sigma \gg g\sqrt{N}$ as can be seen in Fig.~\ref{figTransferRate}.

The disorder-averaged relaxation rate $\overline \gamma$ and the  transport rate $\overline \Gamma$ are related as $\overline \Gamma = \overline \gamma /N$. This relation can be classically understood as a quantum random walk of the excitation, where the dwell time of the excitation on a specific molecule is $1/\overline \gamma$. Consequently, the ratio of the relaxation and  transport rates increases linearly with the number of molecules $N$. The quantum calculation, i.e., the quantum random walk,   shows that the ratio of these two rates is exactly $N$. The  transport rate exhibits a turnover as a function of disorder, which is a consequence of the turnover of the energy-resolved relaxation rate. 

Our calculation thus  explains  the turnover in the transport efficiency numerically observed in Ref.~\cite{Chavez2021}.  However, their numerical investigation  finds an overall scaling of $\propto 1/N^2$ for all parameters of $\sigma$ instead of two scaling behaviors for small-  ($\propto N^{-2}$) and large-disorder ($\propto N^{-1}$)  regimes found in this work. This discrepancy might be a consequence of the average of the logarithmic transport efficiency adopted in Ref.~\cite{Chavez2021}. We note that the resonant  transport rate scales also with $\propto N^{-2}$ for large disorder as shown in Fig.~\ref{figTransferRate}, which is weighted more heavily by an average of the logarithm. Moreover, our findings in Fig.~\ref{figRelaxationRateDisorder}(d) suggest that the disorder-independent regime observed in Ref.~\cite{Chavez2021} is  a specific result of the chosen  parameters rather than a general feature.

The observed turnover is in strict contrast to the Anderson localization, for which the conductivity monotonically decreases with increasing disorder. For charge and exciton  transport in molecules, it is known that noise can lead to a turnover, yet, this is a different mechanism as the disorder enhancement discussed here.

\textbf{Finite-size simulations and thermodynamic limit.} The Green's function solutions in Eqs.~\eqref{eq:greensFktDD} and~\eqref{eq:fullSolutionLaplaceSpaceOld} are exact expressions and thus valid for arbitrary molecule numbers. 
In the derivation of the spectroscopic properties in Sec.~\ref{sec:SingleParticleObservables},  the disorder average of the Green's functions with the Lorentz disorder in Eq.~\eqref{eq:disorderAverage} is formally equivalent to the  thermodynamic limit. Consequently, the expressions for the cavity absorption (i.e., cavity LDOS) and the matter absorption (i.e., bright-state LDOS) are also valid for a finite  molecule number.
In contrast,  the relaxation rate and the transport rate are valid only in the thermodynamic limit $g\rightarrow 0$ and $N\rightarrow \infty$, as we have applied the PPT in the derivation. This can be seen in the finite-size simulation in Figs.~\ref{figRelaxationRateDisorder}(a) and~\ref{figRelaxationRateDisorder}(b), which strongly deviate from the thermodynamic limit  for  low density of states.

\textbf{Implications.} The turnover  as a function of disorder is in strong contrast to the Anderson localization, for which the conductivity is monotonically decaying for increasing disorder.  
 An experimental verification of this turnover  will have a strong technical impact on the design of photovoltaic devices and photo detectors. In order to harness the full spectrum of the sun light, photovoltaic devices  usually work with a broad energy spectrum. This requires that the organic molecules, often deployed in such devices, have either a broad spectral width or a substantial energetic disorder, both have a detrimental influence on the  transport efficiency~\cite{Tessler2009}. An increasing efficiency for larger disorder would thus circumvent this problem. We note that, even though the  article focuses on energy transport, our findings are also valid for charge transport via electron-hole excitations such as in Ref.~\cite{Orgiu2015}.

\subsection{Outlook.}

The Green's function solution in combination with the PPT and the ESM methods is a comprehensive tool, which can be applied to related problems of disordered  ensembles. For example, given the analytical expression for the single-particle Green's function, it is  possible to deal with  nonlinear perturbations. The coherent potential approximation  will be an alternative approach and will be considered elsewhere~\cite{EngelhardtNote,Chenu2017,Chuang2021}.
Due to the absence of  local couplings, the Fano-Anderson model lacks a spatial dimension. In Refs.~\cite{Chavez2021} the total flux is the sum of a cavity-induced contribution and a local-coupling contribution. The latter vanishes exponentially with the system size along with the Anderson localization.  A local coupling term justifies the  random-walk picture on a lattice.
 Further study along this line includes the  multi-mode generalization of the cavity field, dissipation due to the interaction with thermal baths, and  long-range dipolar coupling, i.e.,  F\"orster transport.

\acknowledgements

 G. E. gratefully acknowledges financial support from the China Postdoc Science Foundation (Grant No. 2018M640054), the Natural Science Foundation of China (Grant No. 11950410510  and  No. U1930402) and the Guangdong Provincial Key Laboratory (Grant No.2019B121203002); J. C. acknowledges support from the NSF (Grants No. CHE 1800301 and No. CHE1836913) and the MIT Sloan Fund.

\appendix

\section{Solution of the equation of motion in Laplace space}

\label{sec:DetailsSolutionEoMLaplace}

In this appendix, we derive the Green's function based on  the equations of motion in Laplace space, which is defined by
\begin{equation}
\hat B(z)  = \int_{0}^{\infty} \hat B(t) e^{-zt}dt
\end{equation}
for  operators in the Heisenberg picture $\hat B(t)$. 
Transforming the Heisenberg equation of motion  of the operators $\hat a$ and $\hat B_j$ into Laplace space, we find ~\cite{Engelhardt2016a,Topp2015}
\begin{eqnarray}
	z\hat a(z) - \hat a^{0}  &=& -i E_ C \hat a (z) -i g\sum_{j=1}^N \hat B_j(z), \nonumber \\
	z\hat B_j(z) - \hat B_j^{0}  &=& -i E_ j \hat B_j (z) -i g \hat a (z),
\end{eqnarray}
where $\hat a^{0} = \hat a(0) $ and $ \hat B_j^{0}  = \hat B_j(0)$ denote the Heisenberg operators at time $t=0$.
Due to the tree structure of  the Hamiltonian  in Eq.~\eqref{eq:hamiltonianTerms}, it is possible to directly write  the solution
\begin{eqnarray}
\hat a(z)  &=& \frac{\hat a^{0}}{ z + iE_C} -i  \frac{   g \hat B_j^{0} }{\left( z + i   E_C (z) \right) \left( z + iE_j \right) }, \nonumber     \\
\hat B_j (z)  &=& \frac{\hat B_j^{0}}{ z + iE_j}  - i \frac{g\hat  a^{0} }{ \left( z + i E_j \right)\left( z + i   E_C(z) \right) }  \nonumber  \\
&-& \sum_j  \frac{ g^2\hat B_{j_1}^{0} }{\left( z + i E_j \right)\left( z + i   E_C (z) \right) \left( z + iE_{j_1} \right) }  ,  
\label{eq:solutionHeisenbergOperatorsLaplaceSpace}
\end{eqnarray}
\normalsize
where
\begin{eqnarray}
E_C (z) &=&  E_C  -i  \sum_{j=1}^{N} \frac{g^2}{z + i E_j  }
\end{eqnarray}
is an auxiliary function. After an appropriate multiplication of factors $(z + iE_j) $, the denominators in Eq.~\eqref{eq:solutionHeisenbergOperatorsLaplaceSpace} define the polynomial
\begin{equation}
	\mathcal P(z) =  \left(  z + i  E_C (z)  \right)   \prod_j^{N+1} (z + iE_j),
	\label{eq:DetailsFullCharacteristicPolynomial}
\end{equation}
which is equivalent to the characteristic polynomial of the Hamiltonian in Eq.~\eqref{eq:lightMatterHamiltonian} (up to a replacement of $z\rightarrow - i E$). 
Using the solution in Eq.~\eqref{eq:solutionHeisenbergOperatorsLaplaceSpace}, it is straightforward to construct the Green's function in Laplace space given in Eq.~\eqref{eq:greensFktDD} in the main text. The inverse Laplace transformation is formally defined by
\begin{equation}
	\hat B(t) = \lim_{\delta \rightarrow 0^+} \int_{\delta -i\infty} ^{\delta+i \infty} \hat B(z) e^{zt}dz,
	\label{eq:defInverseLaplaceTransformation}
\end{equation} 
which thus denotes a contour integral  along the imaginary axis. Clearly, this definition is also valid for the inverse Laplace transformation of the Green's functions. If $B(z)$ does not contain any branch cuts, this integral can be evaluated in terms of the residues of $B(z)$. The generalization of  this derivation to the Green's functions of the  transport Hamiltonian in Eq.~\eqref{eq:hamiltonianTranport} is straightforward.

\begin{widetext} 
\section{Detailed derivation of the spectrocopic properties and the local density of states}

In this Appendix, we provide the step-by-step calculations of the spectroscopic properties considered in Sec.~\ref{sec:SingleParticleObservables}.

\subsection{Cavity absorption spectrum}

\label{sec:detailsCavityAbsoporptionSpectrum}

Up to physical constants, the cavity absorption spectrum is equivalent to the cavity LDOS, which reads as, according to the definition in Eq.~\eqref{eq:localDensityOfStates},
\begin{equation}
\nu_{C}  (\omega) = -\frac{1}{\pi}\text{Im} \left[  \frac{i}{\left( z + iE_C(z) \right) } \right] _{z\rightarrow - i\omega}.
\end{equation}
For the Lorentzian disorder, this can be expressed in terms of the eigenvalues of the effective Hamiltonian in Eq.~\eqref{eq:effectiveHamiltonianCBS}, thus
\begin{eqnarray}
\nu_{C}  (\omega) 
&=& - \frac{1}{\pi}\text{Im} \left [ i \frac{z +i E_M +\sigma }{(z + i \epsilon_1)(z + i \epsilon_2) }\right] _{z\rightarrow -i\omega} \nonumber \\
&=& -   \frac{1}{\pi}\text{Im}  \left [ i  \frac{- i \epsilon_1  + i E_M +\sigma }{(i\epsilon_2 - i\epsilon_1)(z + i \epsilon_1) }  +  i\frac{- i \epsilon_2   + i E_M +\sigma }{(i\epsilon_1 -i \epsilon_2)(z +i \epsilon_2) }\right ]  _{z\rightarrow- i\omega}\nonumber  \\
&\equiv&   - \frac{1}{\pi}\text{Im}  \left [A_1^{(C)}  \frac{i}{(z + i  \epsilon_1) }  + A_2^{(C)}  \frac{i}{(z +i   \epsilon_2) } \right ]_{z\rightarrow -i\omega}  , 
\end{eqnarray}
where 
\begin{eqnarray}
A_1^{(C)} &=& \frac{- i \epsilon_1  + i E_M +\sigma }{(i\epsilon_2 - i\epsilon_1) }, \\
A_2^{(C)} &=&  \frac{- i \epsilon_2   + i E_M +\sigma }{(i\epsilon_1 -i \epsilon_2) }.
\end{eqnarray}
This is the form of the cavity LDOS in Eq.~\eqref{eq:CavityLocalDensityOfStates}.

\subsection{Matter absorption spectrum}

\label{sec:molecularAbsorptionSpectrum}

Here, we explicitly evaluate the matter absorption in Eq.~\eqref{eq:molecularAbsorptionGreensFunkt} for the Lorentzian disorder distribution. The step-by-step calculation is
\begin{eqnarray}
\chi_M (\omega) &=& \sum_{i,j}  \text{Im } \left[  G_{i,j}(z )  \right]_{z\rightarrow -i \omega + 0^+}\nonumber \\
&=& \text{Im} \left[  \sum_j   \frac{i }{ z +i E_j }   -\sum_{i,j} \frac{ig^2  }{\left( z +i E_i \right)\left( z +i E_j \right)\left(z + iE_C(z) \right)  } \right]_{z\rightarrow -i \omega + 0^+} \nonumber \\
&=& N  \text{Im} \left[  \int dE    \frac{i }{ z +i E }  P(E)  \right]\nonumber \\
&-& N^2\text{Im} \left[  \int dE_1 dE_2   \frac{ig^2  }{\left( z +i E_1 \right)\left( z +i E_2 \right)\left(z + iE_C(z) \right)  }  P(E_1)P(E_2) \right]_{z\rightarrow -i \omega + 0^+} \nonumber \\
&=& N\text{Im}  \left[ \frac{i }{ z +i (E_M-i\sigma) }    \frac{\sigma}{\pi }  \frac{-2\pi i}{-2\sigma i} \right]_{z\rightarrow -i \omega + 0^+} \nonumber \\
&-& N^2\text{Im} \left[\frac{ig^2  }{\left( z +i(E_M-i\sigma) \right)\left( z +i (E_M-i\sigma)  \right) \left(z + iE_C(z) \right)   }   \left( \frac{\sigma}{\pi }  \frac{-2\pi i}{-2\sigma i} \right)^2  \right]_{z\rightarrow -i \omega + 0^+}  \nonumber \\
&=& N\text{Im} \left[ \frac{i }{ z +i E_M +  \sigma }  \right]_{z\rightarrow -i \omega + 0^+}      - N^2\left[ \text{Im}  \frac{ig^2  }{\left( z +i E_M + \sigma \right) ^2 \left(z + iE_C(z) \right)   }  \right] _{z\rightarrow -i \omega + 0^+}   .
\label{eq:disorderAveragedMolecularAbsorption}
\end{eqnarray}
Next, we show how to express the matter absorption in terms of the Green's function of the cavity mode given in Eq.~\eqref{eq:molecularAbsorption}. To this end, we express the  matter absorption in terms of the eigenvalues of the effective Hamiltonian in Eq.~\eqref{eq:RootsCharPolynomial} as
\begin{eqnarray}
	\chi_M (\omega)  &=& -\frac{N}{\pi}\text{Im}   \left[  \frac{i }{ z +i E_M +  \sigma }       -  \frac{ig^2  N }{\left( z +i E_M + \sigma \right) ^2 \left(z + iE_C(z) \right)   }   \right] _{z\rightarrow -i\omega}\nonumber \\
	&=& -\frac{N}{\pi}\text{Im}   \left[  \frac{i }{ z +i E_M +  \sigma }       -  \frac{ig^2  N }{\left( z +i E_M + \sigma \right)  (z+i \epsilon_1)(z+i \epsilon_2)  }   \right] _{z\rightarrow -i\omega} \nonumber \\
	&=&   - \frac{1}{\pi}N \text{Im}  \left [i  \frac{g^2 N}{\left( -i\epsilon_1 + i E_M + \sigma \right)(i\epsilon_2 - i\epsilon_1)(z +i \epsilon_1) }  + i \frac{g^2 N  }{\left( -i\epsilon_2 +i E_M + \sigma \right) (i\epsilon_1 -i \epsilon_2)(z +i \epsilon_2) }\right ]  _{z\rightarrow -i\omega} \nonumber\\
	&\equiv&    \frac{N}{\pi}\text{Im}  \left [A_1^{(M)}  \frac{i}{(z +i \epsilon_1) }  + A_2^{(M)}  \frac{i}{(z +i \epsilon_2) } \right ]  _{z\rightarrow -i\omega} ,
	\label{eq:molecularAbsorptionDerivation}
\end{eqnarray}
where 
\begin{eqnarray}
A_1^{(M)} &=& -\frac{g^2 N}{\left( -i\epsilon_1 + i E_M + \sigma \right)(i\epsilon_2 - i\epsilon_1) }, \\
A_2^{(M)} &=& - \frac{g^2 N  }{\left( -i\epsilon_2 +i E_M + \sigma \right) (i\epsilon_1 -i \epsilon_2) }.
\end{eqnarray}
The eigenenergies of the effective Hamiltonian \eqref{eq:RootsCharPolynomial} fulfill
\begin{eqnarray}
	\left( \epsilon_1 - E_M +i \sigma \right)\left( \epsilon_2 - E_M +i \sigma \right) = -g^2N ,\\
	\left( \epsilon_2 - E_M + i \sigma \right)\left( \epsilon_1 - E_M +i \sigma \right) = - g^2N,
\end{eqnarray}
from which follows that $A_1^{(M)} = -A_2^{(C)}$ and $A_2^{(M)} = -A_1^{(C)}$. Moreover, we can use the relation
\begin{eqnarray}
	\epsilon_1 +\epsilon_2  = E_C  + E_M -i\sigma
\end{eqnarray}
to interchange the eigenenergies in the last line of \eqref{eq:molecularAbsorptionDerivation}, i.e., 
\begin{eqnarray}
\chi_M (\omega)&\equiv&   - \frac{N}{\pi}\text{Im}  \left [A_2^{(C)}  \frac{i}{(z -i \epsilon_2 + i E_C + iE_M +\sigma ) }  + A_1^{(C)}  \frac{i}{(z - i \epsilon_1 + i E_C + iE_M +\sigma ) } \right ]  _{z\rightarrow -i\omega}  \nonumber \\
  &=&   \frac{N}{\pi}\text{Im}  \left [G_{CC}(-z- i E_C - iE_M - \sigma) \right ]  _{z\rightarrow -i\omega} 
\end{eqnarray}
where we have used Eq.~\eqref{eq:CavityLocalDensityOfStates} for the last step.

\end{widetext}

\section{Analytical techniques }

\label{sec:AnalyticalTechniquesDetails}

Here, we introduce  two  analytical techniques, namely the PPT and the ESM, which are applied in the  calculations in Secs.~\ref{sec:relaxtionDynamics} and \ref{sec:excitationTransfer}.

\subsection{Polynomial perturbation theory}
\label{sec:polynomialPerturbationTheory}

Ordinary time-independent perturbation theory  distinguishes between degenerate and non-degenerate perturbations. PPT unifies both cases by deriving an perturbative expression for the energies based on the characteristic polynomial. 
Let us assume that the characteristic polynomial of a Hamiltonian can be written as a sum of two terms as
\begin{eqnarray}
	\mathcal P (z) &=&  \mathcal P_0  (z)   +  \mathcal  P_1 (z) ,
	\label{eq:fullCharPolynomial}
\end{eqnarray}
where $\mathcal P_0  (z) $ and $\mathcal  P_1 (z)$ denote the unperturbed and  perturbation polynomials, respectively.
We intend to find an approximate expression for the roots   of  $\mathcal P (z)   $, which we write formally as
\begin{equation}
	z_{\mu'}  =  z_{\mu'}^{(0)}+ \delta_{\mu'} ,
\end{equation}
where $z_{\mu'}^{(0)}$ denotes the roots of the  unperturbed polynomial $\mathcal P_0  ( z_{\mu'}^{(0)})=0$ and $\delta_{\mu'}$ is the correction appearing due to $\mathcal  P_1 (z)$. Expanding  Eq.~\eqref{eq:fullCharPolynomial} at $z= z_{\mu'}^{(0)} $ for small $\delta_{\mu'}  $ up to first order we obtain
\begin{eqnarray}
	0 =
	\mathcal  P_1 (  z_{\mu'}^{(0)} ) + \partial_z \mathcal P (z_ {\mu'}^{(0)}) \delta_{\mu'}.
	\label{eq:pptDerivationEquation}
\end{eqnarray}
Resolving Eq.~\eqref{eq:pptDerivationEquation}  for $\delta_{\mu'}$, we readily find the perturbative correction of the roots
\begin{eqnarray}
	\delta_{\mu'}
	&=&  \frac{- \mathcal P  (z_{ \mu'}^{(0)}) }{\partial_z \mathcal P (z_ {\mu'}^{(0)}) } \nonumber \\
	&=& \frac{- \mathcal  P_1 (z_ {\mu'}^{(0)}) }{\prod_{\mu\neq \mu' }  \left(  z_{\mu'}^{(0)} -   z_\mu^{(0)}  \right)    + \partial_z \mathcal  P_1 (z_ {\mu'}^{(0)}) }. \label{eq:polynomialPerturbationTheory}
\end{eqnarray}
This expression interpolates between the degenerate and the non-degenerate perturbation theories. The product in the denominator corresponds to the non-degenerate perturbation theory, while the derivative terms is related to the degenerate perturbation theory.  As the PPT unifies both standard perturbation theories, it is perfectly suitable for the treatment of systems with a continuous spectrum such as reservoirs.

\subsection{Exact stochastic mapping}

\label{sec:ExactStochasticReplacementDetails}

The ESM unravels an analytic  function $ F(z)$ which has no poles in either the lower or upper complex plane in terms of an infinite series of poles, i.e.,
\begin{eqnarray}
	F(z)  =   \lim_{N \rightarrow \infty}  \sum_{ j = 1} ^{N} i \frac{r_j}{z - i E_i } \equiv \lim_{N \rightarrow \infty}  \mathcal F_N(z)
	\label{eq:esrDefinitionIdea}.
\end{eqnarray}
As  $F(z)$ is analytic, the expansion coefficients $r_j$ and the poles $E_j \in \mathbb R$ can be determined by considering    $z =i \omega + 0^+$ with $\omega\in  \mathbb R$ such that
\begin{equation}
	\lim_{N\rightarrow \infty }  \lim_{\delta\rightarrow  0^+} \left|  F(i \omega + \delta)  - \mathcal F_N(i \omega + \delta)  \right|  = 0.
\end{equation}
Using the imaginary part of the  Dirac identity, we find  
\begin{equation}
	r_j  =   \frac{1}{\pi }  \frac{1}{\nu (E_j )}\text{Im}    F(i E_j )  ,
	\label{eq:expansionCoeffients}
\end{equation}
where $\nu(E_j)$ is the density  of poles  defined by
\begin{equation}
	\nu (\omega) =   \lim_{N\rightarrow \infty }  \lim_{\delta\rightarrow 0^+} \frac{1}{\pi}\text{Im} \sum_{ j = 1} ^{N}  i \frac{1}{ i \omega + \delta - i E_i }.
\end{equation}
Even though  using only  the imaginary part of $F(z)$ to define  $r_j$, the real part is  fixed because  of the Kramers-Kronig relations. Since $F(z)$  has no poles in either the upper or lower complex plane, the real and imaginary  parts of analytic functions  are related as
\begin{eqnarray}
	\text{Re}\, F(i\omega )  &=& - \frac{1}{\pi } \int d\omega' \frac{\text{Im} F(i\omega' )}{\omega-\omega' } \nonumber , \\
	\text{Im}\,  F(i\omega )  &=&   \frac{1}{\pi } \int d\omega' \frac{\text{Re} F(i\omega' )}{\omega-\omega' } .
\end{eqnarray}
Using Eq.~\eqref{eq:esrDefinitionIdea} we can show that the Kramers-Kronig relations are  fulfilled by the ESM:
\begin{eqnarray}
	\text{Im}  F(i \omega ) &=&  \lim_{N\rightarrow \infty}  \text{ Re}  \mathcal F_N(i \omega + \delta  )\nonumber \\
	&=&  \lim_{N\rightarrow \infty} \lim_{\delta\downarrow 0} \text{ Re}  \sum_{ j = 1} ^{N} i\frac{r_j}{ i \omega + \delta  - i E_j } \nonumber \\
	&=&  - \lim_{N\rightarrow \infty} \frac{1}{\pi } \sum_{ j = 1} ^{N} \frac{ r_i  }{  (\omega  -  E_j ) }\nonumber  \\
	&=&  -\lim_{N\rightarrow \infty} \frac{1}{\pi } \sum_{ j = 1} ^{N} \text{Im} \frac{   F(i \epsilon_j )  }{\nu (\epsilon_j ) } \frac{ 1  }{  (\omega  -  E_j ) }  \nonumber \\
	&=&  -  \frac{1}{\pi }  \int d\omega'  \frac{ \text{Im}  F(i \omega' )  }{  (\omega  -  \omega'  ) } ,
\end{eqnarray}
demonstrating that the ESM does not lead to any ambiguities related to the definition of the real part of the expansion coefficients $r_j$.
As both $F(z)$ and $\mathcal F_\infty(z)$ are identical on an infinite set of $\mathbb C$, i.e., the imaginary axis, the functions are equal everywhere where defined according to the basic properties of analytical functions.

\section{Details about the finite-size simulation and the fitting  of the relaxation rates}

\label{sec:detailsNumericalEvaluationRates}

We determine the relaxation rate $\gamma(E_1)$ by fitting the Green's function of the donor  with an exponential decaying function, i.e.,
\begin{equation}
	G_{1,1}(t )  = e^{ \left( - i E_1 -\frac{\gamma(E_1) }{2} \right)t }.
	\label{eq:fittingGreensFunction}
\end{equation}
We take advantage of the exact solution of the Green's function in Eq.~\eqref{eq:greensFktDD} and evaluate Eq.~\eqref{eq:fittingGreensFunction} in Laplace space, i.e.,
\begin{equation}
G_{1,1}(z )  = \frac{1}{ z + i E_1 +\frac{\gamma(E_1) }{2} }.
\label{eq:fittingGreensFunctionLaplace}
\end{equation}
Resolving this for the relaxation rate, we find
\begin{equation}
 \gamma(E_1) = -2  \frac{ 1 -\left(z+i E_1 \right)G_{1,1}(z )  }{G_{1,1}(z )} .
\end{equation}
In principle, the right-hand side can be evaluated for arbitrary $z$. For the disorder average, we find that $z = -i E_1 + \delta $ for a small $\delta $ with  $1/ \nu(E_1) \ll\delta\ll \Omega$ converges very fast.

\bibliography{mybibliography}

\end{document}